\documentclass[pdflatex,sn-mathphys-num]{sn-jnl}% Math and Physical Sciences Numbered Reference Style
\usepackage{graphicx}%
\usepackage{multirow}%
\usepackage{amsmath,amssymb,amsfonts}%
\usepackage{amsthm}%
\usepackage{mathrsfs}%
\usepackage[title]{appendix}%
\usepackage{xcolor}%
\usepackage{textcomp}%
\usepackage{manyfoot}%
\usepackage{booktabs}%
\usepackage{algorithm}%
\usepackage{algorithmicx}%
\usepackage{algpseudocode}%
\usepackage{listings}%
\usepackage{amsmath,amssymb,graphicx,amsmath,amssymb,amsthm,physics,bm,bbm,color,xcolor,mathtools,anyfontsize}%
\usepackage[sans]{dsfont}

\usepackage{amsmath}
\usepackage{amsthm}
\usepackage{latexsym}
\usepackage{amssymb}
\usepackage{bm}
\usepackage{bbm}
\usepackage{appendix}
\usepackage{graphics,epstopdf}
\usepackage{physics}
\usepackage{color}
\usepackage[normalem]{ulem}
\usepackage{newlfont}
\usepackage{amsfonts}
\usepackage{amsthm}
\usepackage{graphicx}
\usepackage{epsfig}
\usepackage{mathrsfs}
\usepackage[thinc]{esdiff}
\usepackage{caption}
\usepackage{subcaption}
\newcommand\scalemath[2]{\scalebox{#1}{\mbox{\ensuremath{\displaystyle #2}}}}
\usepackage{times}

\newtheorem{proposition}{Proposition}
\newtheorem{corollary}{Corollary}
\newtheorem{theorem}{Theorem}
\newtheorem{dfn}{Definition}

\newtheorem*{example*}{Example}%
\newtheorem*{remark*}{Remark}%

\newenvironment{prof}{\noindent \textbf{Proof: }\ignorespaces}{\hspace*{\fill}$\Box$\medskip}
\DeclareOldFontCommand{\rm}{\normalfont\rmfamily}{\mathrm}

\def\tr{\operatorname{tr}}
\newcommand \inner[1]{ \left<#1\right>}
\newcommand \floor[1]{ \lfloor#1\rfloor}
\renewcommand \rm[1]{ \mathrm{#1}}
\newcommand \C[1]{ \mathcal{#1}}
\newcommand{\B}{\,\mathcal{B}}
\newtheorem{theoremAlpha}{Theorem}[section]

\newtheorem{propositionAlpha}[theoremAlpha]{Proposition}
\newtheorem{dfnAlpha}[theoremAlpha]{Definition}

\newtheorem*{theorem*}{Theorem}
\newtheorem*{definition*}{Definition}
\newtheorem*{proposition*}{Proposition}
\usepackage[utf8]{inputenc}
\DeclareUnicodeCharacter{2217}{*}

\makeatletter

\@ifundefined{bibinfo}{%
  \newcommand{\bibinfo}[2]{#2}%
}{}

\let\CMP@orig@bibinfo\bibinfo
\renewcommand{\bibinfo}[2]{%
  \def\CMP@key{#1}%
  \def\CMP@volume{volume}%
  \ifx\CMP@key\CMP@volume
  \else
    \CMP@orig@bibinfo{#1}{#2}%
  \fi
}

\@ifundefined{volumename}{}{}
\@ifundefined{bvol}{}{}
\@ifundefined{Bvol}{}{}
\@ifundefined{volume}{}{\renewcommand{\volume}[1]{}}

\makeatother
\theoremstyle{thmstyleone}%
\theoremstyle{thmstyletwo}%

\theoremstyle{thmstylethree}%

\begin{document}

\title[Bochner's theorem for finite inverse semigroups and its connection to Choi's theorem]{Bochner's theorem for finite inverse semigroups and its connection to Choi's theorem}

\author[1]{\fnm{Sohail} \sur{}}\email{sohail.sohail@ttu.edu, sohail.malda@gmail.com}

\author[2,3]{\fnm{Sahil} \sur{}}\email{fsahil@utep.edu, sahil402b2@gmail.com}

\affil[1]{\orgdiv{Department of Computer Science}, \orgname{Texas Tech University}, \orgaddress{\city{Lubbock}, \postcode{79409}, \state{Texas}, \country{USA}}}

\affil[2]{\orgdiv{Department of Physics}, \orgname{University of Texas at El Paso}, \orgaddress{\city{El Paso}, \postcode{79968}, \state{Texas}, \country{USA}}}

\affil[3]{\orgdiv{Optics and Quantum Information Group}, \orgname{The Institute of Mathematical Sciences}, \orgaddress{\city{Chennai}, \postcode{600 113}, \state{Tamil Nadu}, \country{India}}}


\abstract{Bochner's theorem characterizes positive definite functions on groups through the positivity of their Fourier transforms and plays a fundamental role in Harmonic analysis. While Bochner-type results are known for certain classes of semigroups, they typically differ from the group theoretic formulations and do not retain the same level of simplicity and generality. In this work, we prove a Bochner-type theorem for finite inverse semigroups at the level of matrix valued linear maps on the contracted algebras of the semigroups. Using the intrinsic partial order of inverse semigroups, positivity naturally arises through a Möbius-transformed map. Our main result characterizes the positive definiteness of the Möbius transformed map in terms of the positivity of the Fourier transform of the original map with respect to a complete family of inequivalent irreducible representations of the contracted algebra induced by irreducible unitary representations of the maximal subgroups of the inverse semigroup. The proof relies on Fourier inversion formula, Schur orthogonality relations, and alternative characterizations of positive definite maps, all established here in the setting of finite inverse semigroups. As a special case, we show that for the inverse semigroup of matrix units, Bochner's theorem reduces exactly to Choi's characterization of completely positive maps.}


\maketitle
\tableofcontents
\newpage
\section{Introduction}
Positive definite objects and Fourier analysis play a fundamental role in mathematics and physics. In particular, Bochner's theorem provides a complete characterization of positive definite functions on groups in terms of the positivity of their Fourier transforms~\cite{bochner1932,reiter2000classical,rudin1991fourier}. This result underlines large parts of harmonic analysis, representation theory, and probability theory, and has far reaching consequences in both classical and quantum setting~\cite{DAWS20131525,doi:10.1142/S0129167X12501327}.

Bochner-type theorems have also been studied in various semigroup settings, particularly for abelian semigroups with involution and topological semigroups~\cite{Berg1984HarmonicAO}. Although these formulations capture important aspects of positivity, they differ substantially from the group setting, both in their analytic framework and in the role played by representation theory, and do not hold with the same simplicity and generality even for locally compact Hausdorff involution semigroups~\cite{TORBEN_MAACK_BISGAARD}. Bochner-type results have also been obtained in specific scenarios of finite-dimensional conelike semigroups~\cite{Ressel1993}, discrete commutative semigroups with identity and an involution~\cite{lindahl_maserick_1971,berg_maserick_1984}, foundation topological semigroups~\cite{lashkarizadeh_bami_1985}.

Inverse semigroups~\cite{Lawson} provide a natural algebraic framework for describing partial symmetries and arise prominently in the study of groupoids and operator algebraic constructions. Their representation theory is well understood and admits a nice structure via maximal subgroups and induced representations\cite{STEINBERG2006866, STEINBERG20081521}. However, a Bochner-type characterization of positive definite operator-valued maps on finite inverse semigroups does not appear to have been previously developed to the best of our knowledge.

One of the main objectives of this work is to establish a Bochner-type theorem for finite inverse semigroups. More precisely, we formulate and prove a Bochner-type theorem for matrix valued maps on the contracted algebra of a finite inverse semigroup. Due to the intrinsic partial order of inverse semigroups, positivity arises at the level of Möbius transformed map. Our main result characterizes the positive definiteness of the Möbius transformed map in terms of the positivity of the Fourier transform of the original map with respect to a complete family of inequivalent irreducible representations of the contracted algebra induced by a complete family of inequivalent irreducible unitary representations of the maximal subgroups of the inverse semigroup. The proof relies on Fourier inversion formula, Schur orthogonality relations, and alternative characterizations of positive definite maps, all developed here in the setting of finite inverse semigroups.  This framework provides a genuine extension of Bochner's theorem beyond groups, while retaining structural similarities. 
\\

In quantum theory, linear maps between operator algebras encode the dynamics of physical systems. A quantum state is represented by a \emph{density operator}---a positive semidefinite matrix with unit trace. A physical process is then modeled by a linear map that sends density operators to other density operators. However, not every such map corresponds to a valid physical transformation. A key requirement is that the map must be \emph{completely positive} and \emph{trace-preserving}. Such maps are called \textit{quantum channels}. Complete positivity ensures that even when the system is entangled with another (possibly unknown) system, the overall transformation remains valid. This is crucial because in quantum mechanics, systems are often not isolated.  A foundational result that provides a practical criterion for complete positivity is the \textit{Choi theorem} \cite{CHOI1975285}. It states that a linear map is completely positive if and only if its associated \emph{Choi matrix} is positive semidefinite. Choi's theorem and the Choi matrix are essential tools for the mathematical analysis of quantum channels and superchannels~\cite{wolf_cirac_2008,chruscinski_kossakowski_2009,SohailAHP2025,Darek_2026,Sohail_2021PLA,das2024,Johnston_2011, Chiribella, Gour}.
\\

At first sight, Bochner's theorem and Choi's theorem arise in rather different settings: the former concerns positive definite functions or maps on groups and their Fourier transforms, while the latter provides matrix positivity criterion for completely positivity of linear maps between operator algebras. Nonetheless, both results are, from a broader perspective, statements relating positivity in an ``original domain" to positivity in a ``transformed domain". This observation leads to the following question: \textit{Can Choi's theorem on complete positivity be understood as a special case of Bochner-type theorem?} Our formulation of the Bochner-type theorem answers this question in affirmative. In particular, this serves as an important and illuminating special case of our general results when the inverse semigroup of matrix units is considered. The full matrix algebra over the complex field is a contracted semigroup algebra of the inverse semigroup of matrix units~\cite{Amitsur_1951}. In this setting, the Fourier transform of a linear map on the matrix algebra coincides exactly with the Choi matrix of the map, and the Bochner-type theorem reduces to Choi's theorem on complete positivity. From this perspective, Choi's theorem emerges as a special instance of a more general Bochner-type result, rather than as an isolated criterion.

\subsection{Our Contributions and Overview of the Main Results}
The main goal of this work is to establish a Bochner's theorem for finite inverse semigroups and to recover Choi's theorem on complete positivity as a special case. We begin by defining a convolution product between maps from the contracted semigroup algebra $\mathbb{C}_0[S]$ of a semigroup $S$ to an arbitrary associative algebra $\mathcal{A}$. The convolution product makes the space $L(\mathbb{C}_0[S],\mathcal{A})$ of linear maps from $\mathbb{C}_0[S]$ to $\mathcal{A}$ an associative algebra.  Then we show that the convolution algebra $L(\mathbb{C}_0[S],\mathcal{A})$ and the tensor product algebra $\mathbb{C}_0[S] \otimes \mathcal{A}$ are isomorphic [see Theorem~\ref{thm:semigroup convolution theorem}]. As a consequence of Theorem~\ref{thm:semigroup convolution theorem}, in the specific case of the matrix unit semigroup, we identify the product in the space of maps on the matrix algebras that is preserved by the Choi-Jamiołkowski isomorphism as convolution. Then, defining the Fourier transform of a map from $\mathbb{C}_0[S]$ to $M_n(\mathbb{C})$, we derive a Fourier inversion formula when $S$ is a finite inverse semigroup [ see Theorem~\ref{thm:fourier_inversion_inverse_semigroup}], which is one of our main results. As a corollary of Theorem~\ref{thm:fourier_inversion_inverse_semigroup}, we show that for the inverse semigroup of matrix units, the Fourier transform of a map with respect to the identity representation coincides with the Choi matrix of the map, and the Fourier inversion formula reduces to the Choi inversion formula [see Corollary~\ref{corollary:Fourier inversion becomes Choi inversion}]. Additionally, We derive a Plancherel formula in the finite inverse semigroup setting [see Theorem~\ref{thm:Placheral formula for inverse semigroup}]. 

Then, by introducing the notion of matrix valued positive definite maps on contracted semigroup algebras, we prove a Bochner's theorem for finite inverse semigroups in Theorem~\ref{thm:Bochner's theorem for finite inverse semigroups}, which is the central result of this work. The proof relies on Fourier inversion formula [see Theorem~\ref{thm:fourier_inversion_inverse_semigroup}], Schur orthogonality relations [see Proposition~\ref{Schur Orthogonality relation for inverse semigroup}], and alternative characterizations of positive definite maps [see Proposition~\ref{prop:alternative characterization of positive definite maps 1}, and Proposition~\ref{prop:alternative characterization of positive definite maps 2}], all developed here in the setting of finite inverse semigroups. Then, in Section~\ref{subsec:Choi theorem on Completely positive maps and Bochner's theorem on positive definite maps on finite dimensional matrix algebra}, we show that Bochner's theorem reduces to Choi theorem on completely positive maps when inverse semigroup of matrix units are considered.
Finally, in Theorem~\ref{thm:CP vs positivity}, we derive a necessary and sufficient condition on a representation  $\rho:M_m(\mathbb{C}) \to M_{d_{\rho}}(\mathbb{C})$ such that the Complete positivity vs. positivity correspondence holds between a linear map $\Phi: M_m(\mathbb{C}) \to M_n(\mathbb{C})$  and its Fourier Transform $\widehat{\Phi}(\rho)$.
\subsection{Structure of the paper}
The structure of the paper is as follows.
In Section~\ref{Sec:Preliminaries}, we review basic notions of semigroups, inverse semigroups, their contracted algebras, and the corresponding representation theory. Section~\ref{sec:Main Results} contains the main results of the paper and is divided into several subsections. In Section~\ref{subsec:Fourier transform of maps}, we define the Fourier transform of maps on the contracted algebra of a finite inverse semigroup and discuss its various properties.  In Section~\ref{subsec:Fourier Inversion for finite inverse semigroup}, we derive a Fourier inversion formula for finite inverse semigroups and show that in the case of inverse semigroup of matrix units, this formula reduces to the Choi inversion formula. A Plancherel formula is also derived in the finite inverse semigroup setting. In Section~\ref{subsec:Positive definite maps and Bochner's theorem for finite inverse semigroup}, we prove a Bochner's theorem for finite inverse semigroups. Two alternative characterizations of positive definite maps on contracted algebras of finite inverse semigroups are provided, and a Schur orthogonality relation in the setting of finite inverse semigroups is proved, which plays a crucial role in the proof of Bochner's theorem. Section~\ref{subsec:Choi theorem on Completely positive maps and Bochner's theorem on positive definite maps on finite dimensional matrix algebra} is devoted to the finite dimensional matrix algebra case, where we show that Bochner's theorem reduces to Choi's characterization of completely positive maps. In Section~\ref{subsec:Relation between complete positivity of maps and positivity of their Fourier transforms}, we derive a necessary and sufficient condition on a representation $\rho:M_m(\mathbb{C}) \to M_{d_{\rho}}(\mathbb{C})$ such that the Complete positivity vs. positivity correspondence holds between a linear map $\Phi: M_m(\mathbb{C}) \to M_n(\mathbb{C})$  and its Fourier Transform $\widehat{\Phi}(\rho)$. Finally, Section~\ref{sec:Conclusion} contains concluding remarks and perspectives.
\section{Preliminaries} \label{Sec:Preliminaries} 
In this section, we provide a brief overview of the fundamental concepts of finite inverse semigroups along with their key properties and representations. We also present a concise discussion of Choi's theorem on completely positive maps.
\subsection{Semigroup, Inverse Semigroup and their Algebras: Structure and Representations}
 A semigroup is a pair $(S, \cdot)$ of a non-empty set $S$ and a binary operation $\cdot : S \times S \to S$ which is associative, \emph{i.e.,} $x\cdot(y \cdot z)=( x\cdot y) \cdot z$ for all $x,y,z \in S$. Throughout the paper, we adopt the notation $x \cdot y=xy$.
\begin{definition*}[Semigroup Algebra~\cite{Clifford_semigroup}]
    Let $S$ be a semigroup. A semigroup algebra of $S$ over the complex field $\mathbb{C}$ is an algebra $\mathbb{C}[S]$ with a basis which is isomorphic to the semigroup $S$.
\end{definition*}
 By the abuse of notation, we use $S$ itself to denote the basis for $\mathbb{C}[S]$.
\begin{definition*}[Contracted Algebra~\cite{Clifford_semigroup}]
    A contracted algebra $\mathbb{C}_{0}[S]$ of a semigroup $S$ is an algebra over the complex field $\mathbb{C}$ with a basis $\mathscr{B}$ such that $\mathscr{B} \cup \{ 0\}$ is isomorphic to $S$, where $0$ denotes the zero of the contracted algebra.
\end{definition*}
By the abuse of notation, we use $S\setminus\{0\}$ (or sometimes $S \setminus \{z\}$) to denote a basis of $\mathbb{C}_0[S]$. In the following, we provide a brief description of the basic concepts and properties of an inverse semigroup. Unlike a group, an inverse semigroup generalizes the notion of inverse without requiring the presence of an identity element.
\begin{dfnAlpha}
    [Inverse Semigroup~\cite{Clifford_semigroup, Lawson}]\label{dfn:inverse semigroup}
    A semigroup $S$ is called an inverse semigroup if, for every element $s \in S$, there exists a unique element $s^{-1}$ such that $ss^{-1}s=s$ and $s^{-1}ss^{-1}=s^{-1}$.
\end{dfnAlpha}
It immediately follows from the definition that $(s^{-1})^{-1}=s$. An element $p \in S$ is called idempotent if $p^{2}=p$. We now list some elementary properties that hold in an inverse semigroup.
\begin{proposition*}[\cite{Clifford_semigroup,Lawson}]
    Let $S$ be an inverse semigroup. Then the following properties hold.
    \begin{enumerate}
        \item For any element $s \in S$, the elements $ss^{-1}$ and $s^{-1}s$ are idempotent.
        \item If $e \in S$ is an idempotent, then $e^{-1}=e$.
        \item Let $e$ and $f$ be two idempotents in $S$. Then $e f$ is also idempotent, and $e  f=f e$, i.e., the idempotents commute.
        \item For $s,t \in S$, we have $(st)^{-1}=t^{-1}s^{-1}$.
    \end{enumerate}
\end{proposition*}
\subsubsection{Partial Ordering and Möbius Inversion}

Let $S$ be a finite partially ordered set and $R$ a commutative unital ring. The incidence algebra of $S$ over $R$, denoted by $R[S \times S]$, is an algebra of all the functions $f: S \times S \to R$ such that
\begin{align}
    f(x,y)=0
\end{align}
whenever $x,y \in S$ are incomparable \emph{i.e.,} $x \nleq y$. The multiplication between $f,h \in R[S \times S]$ is given by the convolution defined as follows~\cite{Stanley}:
\begin{align}
    f * h(x,y):=\sum_{\substack{z \in S \\ x \leq z \leq y} } f(x,z) h(z,y).
\end{align}

The incidence algebra is a unital algebra with the identity element given by $\mathrm{I}(x,y)=\delta_{xy}$, where $\delta_{xy}$ denotes the Kronecker delta. With the identity being present in the incidence algebra, the inverse of $f \in R[S \times S]$ can be defined if there exist $f^{-1} \in R[S \times S]$ such $f*f^{-1}(x,y)=f^{-1}*f(x,y)=\rm{I}(x,y)$ for all $x,y \in S$ with $x \leq y$. Equivalently,
\begin{align}
   \sum_{\substack{z \in S \\ x \leq z \leq y} }\!\! f(x,z) f^{-1}(z,y) =\delta_{x,y}. \label{eq:mobius inversion_1}
\end{align}
The above equation is used to find $f^{-1}(x,y)$ inductively. Specifically, for all $x \in S$, we have $f^{-1}(x,x)=\frac{1}{f(x,x)}$. Then using this value one can compute $f^{-1}(x,z)$ for those elements $z \in S$ such that there is no $z_0 \in S$ with $x < z_0 < z$. Proceeding inductively in this way, one can compute $f^{-1}(x,y)$.
\begin{definition*}[Zeta function and Möbius function~\cite{Stanley}]
    The zeta function is the element $\zeta \in R[S \times S]$ defined by 
    \[
        \zeta(x,y)=
        \begin{cases}
            1, & \text{if} \quad x \leq y,\\
            0,  & \text{otherwise.}
        \end{cases}
    \]
    The Möbius function, denoted by $\mu$, is the inverse of the zeta function in the incidence algebra.
\end{definition*}
It is clear that $\mu(x,x)=\frac{1}{\zeta(x,x)}=1$. Moreover, when $x <y$, from Eq.~(\ref{eq:mobius inversion_1}) we have 
\begin{align}
    &\sum_{\substack{z \in S \\ x \leq z \leq y} } \!\!\zeta(x,z)\mu(z,y) =0, \nonumber \\
    & \iff \sum_{\substack{z \in S \\ x \leq z \leq y} } \!\!\mu(z,y) =0, \nonumber \\
 &\iff   \mu(x,y)=-\!\!\!\! \sum_{\substack{z \in S \\ x < z \leq y} } \!\!\mu(z,y). \label{eq:Mobius function inductively}
\end{align}
Equation~(\ref{eq:Mobius function inductively}) provides a way to inductively determine the Möbius function. 

\begin{theoremAlpha}[Möbius inversion formula~\cite{Stanley}]
    Let $S$ be a partially ordered set with the ordering relation $\leq$, and let $R$ be a field. Suppose $f,h: S \to R $ are functions satisfying, for all $x \in S$,
    \begin{align}
        f(x)=\sum_{\substack{y \in S \\ y \leq x}} h(y).
    \end{align}
Then $h$ can be expressed in terms of $f$ as 
\begin{align}
     h(x)=\sum_{\substack{y \in S \\ y \leq x}} \mu(y,x)f(y).
\end{align}
\end{theoremAlpha} 
\subsubsection{Steinberg's Isomorphism and Representations of the Contracted Algebra of Finite Inverse Semigroups}
An example of an inverse semigroup is the symmetric inverse semigroup. Denoted by $\mathcal{J}_{X}$, it consists of all partial bijections $\tau: Y \to Z$ on a non-empty set $X$, where $Y,Z \subseteq X$. The binary operation is the composition of partial transformations. It can be verified that $\tau: Y \to Z$ and its inverse $\tau^{-1}: Z \to Y$ satisfy all the defining properties of an abstract inverse semigroup as given in Definition~\ref{dfn:inverse semigroup}. The Preston-Vagner representation theorem states that any inverse semigroup $S$ can be thought of as an inverse subsemigroup of $\mathcal{J}_{S}$.
\begin{theorem*}[Preston-Vagner representation~\cite{Clifford_semigroup, Lawson, lawson2023_Inverse_Semigroup}]
    Let $S$ be an inverse semigroup, and let $\mathcal{J}_S$ denote the symmetric inverse semigroup of all partial bijection of $S$. Then $S$ is isomorphic to an inverse subsemigroup of $\mathcal{J}_S$.
\end{theorem*}
For $\tau \in \mathcal{J}_S$, denote by $\rm{dom}(\tau)$ the domain of $\tau$ and by $\rm{ran}(\tau)$ its range. Denoting by $\rm{id}_{U}$ the identity map on a subset $U \subseteq S$, we have $\tau \tau^{-1}= \rm{id}_{\rm{ran}(\tau)}$ and $\tau^{-1} \tau= \rm{id}_{\rm{dom}(\tau)}$. Since $\rm{id}_U$ can be identified with the subset $U \subseteq S$ itself, it is customary to define, for an abstract inverse semigroup $S$ and $s \in S$,
\begin{align}
    \mathrm{dom}(s):=s^{-1}s, \quad \mathrm{ran}(s):=ss^{-1},
\end{align}
and to regard $s$ as an isomorphism $s:s^{-1}s \to ss^{-1}$. Recalling that $s^{-1}s$ and $ss^{-1}$ are idempotents, the notion of two idempotents being isomorphic is defined as follows.
\begin{definition*}[Isomorphic idempotents~\cite{STEINBERG20081521}]
    Let $S$ be an inverse semigroup. Two idempotents $e,e' \in S$ are said to be isomorphic if there exists an element $s \in S$ such that 
    \[e=s^{-1}s\quad \text{and}\quad e'=ss^{-1}.\]
\end{definition*}
 A partial ordering on an inverse semigroup $S$ is defined as follows.
\begin{dfnAlpha}[Partial order~\cite{Lawson,lawson2023_Inverse_Semigroup}] \label{dfn:partil_ordering_in_inverse_semigroup}
    Let $S$ be an inverse semigroup and let $s,t \in S$. We define $s \leq t$ if there exist an idempotent $e \in S$ such that $s= et$.
\end{dfnAlpha} 
For two arbitrary elements $s,t \in S$ of an inverse semigroup $S$, the following statements are equivalent:
\[\begin{aligned}
    & \text{(i)} && s \leq t,\\
        & \text{(ii)} && \text{there exists and idempotent } e' \text{ such that } s=te',\\
        & \text{(iii)} && s=ss^{-1}t, \\
        & \text{(iv)} && s=ts^{-1}s.
\end{aligned}\]  
\begin{dfnAlpha}[Groupoid basis~\cite{STEINBERG20081521} ]\label{dfn:groupoid_basis}
    Let $S$ be a finite inverse semigroup. For every non-zero element $s \in S$, define
    \begin{align}
        \lfloor s \rfloor := \sum_{\substack{t \in S\\ t \leq s}} \mu(t,s) t \in \mathbb{C}_{0}[S], \label{eq:groupoid_basis}
    \end{align}
    where $\mu(t,s)$ is the M\"obius function defined on the natural partial order on $S$, as given in Eq.~(\ref{eq:Mobius function inductively}).
\end{dfnAlpha}
It immediately follows that for the zero element $z \in S$, $\floor{z}=0$.
It is proved in Ref.~\cite{STEINBERG20081521} that $\{ \lfloor s \rfloor \ :s \in S\setminus \{z\} \}$ forms a basis for the contracted algebra $\mathbb{C}_{0}[S]$. The basis $\{\lfloor s \rfloor\}_{s \in S \setminus \{z\}}$ is known as the groupoid basis. Using the M\"obius inversion formula, the natural basis $\{s\}_{s \in S \setminus \{z\}}$ can be recovered as follows:
\begin{align}
    s= \sum_{\substack{t \in S\\ t \leq s}} \lfloor t \rfloor.
\end{align}
The basis elements in $\{\lfloor s \rfloor\}_{s \in S \setminus \{z\}}$ multiply as follows \cite{STEINBERG20081521}:
\begin{align}
    \lfloor s \rfloor \lfloor t \rfloor = 
       \begin{cases}
                \lfloor st \rfloor  &  \text{if}\ s^{-1}s =tt^{-1}, \\ 
                0 & \text{otherwise}.
       \end{cases}\label{gropoid_basis_multiplication}
 \end{align}
 \begin{definition*}[Green's $\mathcal{D}$-relation~\cite{Clifford_semigroup, Lawson, STEINBERG20081521}]
     Let $S$ be a finite inverse semigroup. For $s,t \in S$, we say that $s$ and $t$ are  $\mathcal{D}$-related, written $s \mathcal{D}t$, if there exists an element $x \in S$ such that $x^{-1}x = ss^{-1}$ and $xx^{-1}= tt^{-1}$.
 \end{definition*}
 The relation defined above is called Green's $\mathcal{D}$-relation. It can be shown that this relation is an equivalence relation on $S$ and its equivalence classes are referred to as the $\mathcal{D}$-classes. Let $\{ D_k \}_{k=1}^n$ denote the $\mathcal{D}$-classes of $S$. Define
  \[\mathbb{C} D_k := \mathrm{Span} \{ \lfloor s \rfloor : s \in D_k\}.\] It should be noted that there always exists an idempotent inside each $\mathcal{D}$-classes $D_k$. Indeed for any $t \in D_k$, one has $t^{-1}t \in D_k$. Moreover, if $e$ and $f$ are two idempotents in $D_k$, then they are isomorphic, \emph{i.e.,} there exists $x \in S$ such that $x^{-1}x=e$ and $xx^{-1}=f$. 
 \begin{propositionAlpha} \label{Prop:CD-algebra}
    For each $\mathcal{D}$-class $D_k$, the vector space $\mathbb{C} D_k$ is a subalgebra of $\mathbb{C}_0[S]$.
 \end{propositionAlpha}
 \begin{prof}
     From the definition, $ \mathbb{C} D_k$ is a vector space. Let $s,t \in D_k$ be two arbitrary elements. So, the multiplication of $\floor{s}$ and $\floor{t}$ is given by Eq.~(\ref{gropoid_basis_multiplication}). It is sufficient to show that $st \in D_k$. Since $s,t \in D_k$, there exist $x \in S$, such that $x^{-1}x=ss^{-1}$ and $xx^{-1}=tt^{-1}$. Now, 
     \begin{eqnarray*}
         (st)(st)^{-1}&=stt^{-1}s^{-1}=ss^{-1}ss^{-1}=ss^{-1}=x^{-1}x,
     \end{eqnarray*}
    where in the 2nd equality, we have used the condition in Eq.~(\ref{gropoid_basis_multiplication}). It follows from the above equation that $(st) \mathcal{D} t$ and hence $st \in D_k$. This completes the proof.
 \end{prof}
 \\
 
 A subset $G \subseteq S$ is called a subgroup of a semigroup $S$ if $G$ forms a group under the operation in $S$. A subgroup $G$ of $S$ is called a maximal subgroup if it is not properly contained in any other subgroup of $S$.
 \begin{proposition*}[\cite{STEINBERG20081521}]
      Let $S$ be a finite inverse semigroup, and let $e \in S$ be an idempotent. Then $e$ is the identity element of a unique maximal subgroup $G_e \subseteq S$ given by 
     \begin{align*}
         G_e= \{s \in S : ss^{-1}=s^{-1}s=e\}.
     \end{align*}
 \end{proposition*}
 For an idempotent $e \in S$, the subgroup $G_e$ is called the \emph{maximal subgroup at $e$}. The following proposition states that maximal subgroups at different idempotents belonging to the same $\mathcal{D}$-class are isomorphic.
 \begin{proposition*}[\cite{STEINBERG2006866,STEINBERG20081521}]
      Let $e$ and $f$ be two isomorphic idempotents of a finite inverse semigroup $S$, and let $G_e$ and $G_f$ denote their maximal subgroups. Then $G_e$ is isomorphic to $G_f$.  
 \end{proposition*}
 
 From Proposition~\ref{Prop:CD-algebra} and the multiplication rule given in Eq.~(\ref{gropoid_basis_multiplication}), it follows that 
 \[\mathbb{C}_{0}[S]= \bigoplus_{k=1}^n \mathbb{C} D_k.\] For each $D_k$, let us fix an idempotent $e_k$. Let $G_{e_k}$ denotes the maximal subgroup of $S$ at $e_k$. For each idempotent $e \in D_k$, let us fix an element $p_e \in S$ such that $\mathrm{dom}(p_e)=e_k$ and $\mathrm{ran}(p_e)=e$ with $p_{e_k}=e_k$. These choices will be fixed throughout the remainder of the paper. The following theorem states that $\mathbb{C}D_k$ is isomorphic to $M_{r_k}(\mathbb{C}G_{e_k})$, where $M_{r_k}(\mathbb{C}G_{e_k})$ is the algebra of $r_k \times r_k$ matrices over the group algebra $\mathbb{C} G_{e_k}$ with $r_k$ being the number of idempotents in $D_k$. 
 \begin{theoremAlpha}[\cite{STEINBERG20081521}] Let $S$ be a finite inverse semigroup, and let $\{ D_k\}_{k=1}^N$ denote its $\mathcal{D}$-classes. For each $k$, fix an idempotent $e_k \in D_k$, and let $G_{e_k}$ be the maximal subgroup of $S$ at $e_k$. Then \[ \mathbb{C}_{0}[S] \cong \bigoplus_{k=1}^N M_{r_k}(\mathbb{C}G_k).\] Specifically,
     The algebra $\mathbb{C}D_k$ is isomorphic to the algebra $M_{r_k}(\mathbb{C}G_k)$, where the isomorphism is given by 
     \begin{align}
         \varphi (\lfloor s \rfloor)= p^{-1}_{\mathrm{ran}(s)} s~p_{\mathrm{dom}(s)} E_{\mathrm{ran}(s), \mathrm{dom}(s)},
     \end{align}
     and the inverse map is given by
     \begin{align}
         \varphi^{-1} (s E_{t,f})= \lfloor p_t s p^{-1}_f \rfloor,
     \end{align}
     where $E_{a,b}$ is the $r_k \times r_k$ matrix indexed by the idempotents of $D_k$ whose $a,b$-th entry is $1$ and the rest of the entries are zero. 
 \end{theoremAlpha}
 \begin{propositionAlpha} \label{prop:D_k to G_(e_k)}
     Let $s \in D_k$. Then $p^{-1}_{\rm{ran}(s)} s~ p_{\rm{dom}(s)}  \in G_{e_k}$.
 \end{propositionAlpha}

 \begin{prof}
  We begin by evaluating $\left(p^{-1}_{\mathrm{ran}(s)} s p_{\rm{dom}(s)} \right)^{-1}  \left(p^{-1}_{\rm{ran}(s)} s p_{\rm{dom}(s)} \right)$.
  \begin{align*}
      \left(p^{-1}_{\rm{ran}(s)} s p_{\rm{dom}(s)} \right)^{-1}  \left(p^{-1}_{\rm{ran}(s)} s p_{\rm{dom}(s)} \right)&= p^{-1}_{\rm{dom}(s)} s^{-1} p_{\rm{ran}(s)} p^{-1}_{\rm{ran}(s)} s p_{\rm{dom}(s)} \\
      &= p^{-1}_{\rm{dom}(s)} s^{-1} \rm{ran}(s) s p_{\rm{dom}(s)} \\
      &= p^{-1}_{\rm{dom}(s)} s^{-1} ss^{-1} s p_{\rm{dom}(s)} \\
      &= p^{-1}_{\rm{dom}(s)} s^{-1} s p_{\rm{dom}(s)}\\
      &= p^{-1}_{\rm{dom}(s)} \rm{dom}(s) p_{\rm{dom}(s)}\\
      &= p^{-1}_{\rm{dom}(s)} p_{\rm{dom}(s)} p^{-1}_{\rm{dom}(s)} p_{\rm{dom}(s)}\\
      &= p^{-1}_{\rm{dom}(s)} p_{\rm{dom}(s)}\\
      &= e_k.\\
  \end{align*}
  In a similar computation shows that \[ \left(p^{-1}_{\rm{ran}(s)} s p_{\rm{dom}(s)} \right)  \left(p^{-1}_{\rm{ran}(s)} s p_{\rm{dom}(s)} \right)^{-1}=e_k.\] Thus $p^{-1}_{\rm{ran}(s)} s~ p_{\rm{dom}(s)}$ is an element of the maximal subgroup $G_{e_k}$. This completes the proof.
 \end{prof}
 \\
 \begin{propositionAlpha} \label{prop:Transformation D_k to G_{e_k}}
     For all $u \in G_{e_k}$ and idempotents $a,b \in D_k$, we have 
     \[\rm{ran}(p_a \, u \, p^{-1}_b)=a \quad \text{and} \quad \rm{dom}(p_a \, u \, p^{-1}_b)=b.\]
 \end{propositionAlpha}
 \begin{prof}
 We compute:
     \begin{align*}
         \rm{ran}(p_a \, u \, p^{-1}_b)&=p_a \, u \, p^{-1}_b\, p_b\, u^{-1}\, p^{-1}_a \\
         &= p_a \, u \, e_k\, u^{-1}\, p^{-1}_a \\
          &= p_a \, u \, u^{-1}\, p^{-1}_a \\
          &= p_a \,e_k\, p^{-1}_a \\
          &= p_a \,p^{-1}_a \,p_a\, p^{-1}_a \\
          &=p_a \,p^{-1}_a \\
          &=a.
     \end{align*}
   A similar computation shows that \[\rm{dom}(p_a \, u \, p^{-1}_b)=b.\] This completes the proof. 
 \end{prof}
 \begin{propositionAlpha}\label{$s=t$ is the only option}
    Let $S$ be a finite inverse semigroup and let  $s,t \in S$ be nonzero elements satisfying $ss^{-1}=tt^{-1}$. Then $s^{-1}t$ is an idempotent if and only if $s=t$.
\end{propositionAlpha}
\begin{prof}
    When $s=t$, trivially $s^{-1}t$ is an idempotent. Conversely, suppose that $s^{-1}t$ is an idempotent. Then
    \begin{align}
       s^{-1}ts^{-1}t=s^{-1}t, \label{SS4} 
    \end{align}
    \begin{align}
    s^{-1}t=t^{-1}s, \label{SS5}
    \end{align}
    and 
    \begin{align}
        (s^{-1}t)^{-1}(s^{-1}t)=(s^{-1}t)(s^{-1}t)^{-1}. \label{SS6}
    \end{align}
    Using Eq.~(\ref{SS6}) and the assumption $ss^{-1}=tt^{-1}$, we have 
    \begin{align}
        s^{-1}s=t^{-1}t. \label{SS9}
    \end{align}
    Now we use Eq.~(\ref{SS5}) in Eq.~(\ref{SS4}) to obtain
    $$s^{-1}tt^{-1}s=s^{-1}t 
       \iff s^{-1}ss^{-1}s=s^{-1}t 
       \iff s^{-1}s=s^{-1}t 
       \iff ts^{-1}s=ts^{-1}t.$$ Now, using Eq.~(\ref{SS9}), we arrive at
    \begin{align}
        tt^{-1}t=ts^{-1}t, \label{SS7} 
    \end{align}
    which is equivalent to 
    \begin{align}
        t=ts^{-1}t. \label{SS8}
    \end{align}
     Since in a finite inverse semigroup every element has a unique inverse, we conclude from Eq.~(\ref{SS8}) that $s=t$. This completes the proof.
\end{prof}
 \\
 
 Steinberg's isomorphism provides a method for constructing a complete set of inequivalent irreducible representations of the contracted algebra $\mathbb{C}_{0}[S]$ from complete sets of inequivalent irreducible representations of the maximal subgroups of a finite inverse semigroup $S$.
 \begin{theoremAlpha}[\cite{STEINBERG20081521}] \label{lemma:Stineberg_1}
     Let $S$ be a finite inverse semigroup, and let $\{ D_k\}_{k=1}^N$ denote its $\mathcal{D}$-classes. For each $k$, fix an idempotent $e_k \in D_k$, and let $G_{e_k}$ be the maximal subgroup of $S$ at $e_k$. Let $\mathrm{Irr}(G_{e_k})$ be a complete set of inequivalent irreducible representations of the maximal subgroup $G_{e_k}$. For $\rho_k \in \mathrm{Irr} (G_{e_k})$, define a representation $\overline{\rho}_k$ of $M_{r_k}(\mathbb{C}G_{e_k})$ by 
     \[\overline{\rho}_k(g E_{a,b}):= E_{a,b} \otimes \rho_k (g),\] extended linearly to the whole algebra. Now define $\overline{\rho}_k$ on the algebra $\bigoplus_{k=1}^N M_{r_k}(\mathbb{C}G_{e_k})$ by defining its action on the summands other than $k$ as zero. Then the set of all such $\overline{\rho}_k$ forms a complete set of inequivalent irreducible representations of the contracted semigroup algebra $\mathbb{C}_{0}[S]$. Specifically, the representation $\overline{\rho}_k$ is given on the groupoid basis by
 \begin{align}
     \overline{\rho}_k(\floor{s})=\begin{cases}
                                   E_{\rm{ran}(s),\rm{dom}(s)} \otimes \rho_{k}(p^{-1}_{\rm{ran}(s)}\, s \, p_{\rm{dom}(s)}) & \text{if} \quad s \in D_k, \\
                                   0 & \text{otherwise}.
                                  \end{cases}
 \end{align}
 \end{theoremAlpha}
\subsection{Choi-Jamiołkowski Isomorphism and Completely Positive Maps} \label{s2}
Let
\(\mathcal{H}_1\!=\!\mathbb{C}^n\) and \(\mathcal{H}_2 = \mathbb{C}^m\) be two finite-dimensional Hilbert spaces, and let $\mathcal{B}(\mathcal{H}_1)$ and $\mathcal{B}(\mathcal{H}_2)$  respectively denote the algebras of bounded linear operators on them. Let $ \{e_{ij}\}_{i,j=1}^n$ and $ \{f_{ij}\}_{i,j=1}^m$ respectively denote the complete set of matrix units for $\mathcal{B}(\mathcal{H}_1)$ and $\mathcal{B}(\mathcal{H}_2)$. Given a linear map $\Phi :\mathcal{B}(\mathcal{H}_1)\rightarrow \mathcal{B}(\mathcal{H}_2)$, its Choi matrix \cite{book,article,CHOI1975285,kraus1983states,Sudarshan1985} is defined by
\begin{eqnarray}
        C_{\Phi} =\sum_{{i,j}=1}^{n} e_{ij} \otimes \Phi(e_{ij}) \in \mathcal{B}(\mathcal{H}_1)\otimes \mathcal{B}(\mathcal{H}_2). \label{CJKSdefn}
\end{eqnarray} 
The map $\Phi \mapsto C_{\Phi}$ is known as the Choi-Jamiołkowski isomorphism. Its inverse is given by
\begin{eqnarray}
      \Phi(A)=\tr_{1}[(A^{\tau} \otimes \mathbb{I})C_{\Phi}], \label{eq:Choi_inversion}
  \end{eqnarray}
  where $\tau$ denotes transposition with respect to the standard basis of $\mathcal{H}_1$, and $\mathbb{I}$ denotes the identity operator on $\mathcal{H}_2$. Equation~\ref{eq:Choi_inversion} will be referred to as the Choi inversion formula throughout the paper. 

  A linear map $\Phi : \mathcal{B}(\mathcal{H}_1) \rightarrow \mathcal{B}(\mathcal{H}_2)$ is called positive if it maps positive elements of $\mathcal{B}(\mathcal{H}_1)$ to positive elements of $\mathcal{B}(\mathcal{H}_2)$. Given a linear map $\Phi$, one can construct a map
  \[\mathrm{id}_{k} \otimes\Phi:M_k(\mathbb{C}) \otimes \mathcal{B}(\mathcal{H}_1) \rightarrow M_k(\mathbb{C}) \otimes \mathcal{B}(\mathcal{H}_2),\] where $M_k(\mathbb{C})$ denotes the algebra of $k \times k$ complex matrices.
\begin{dfnAlpha} \label{D_333}
    A linear map $\Phi : \mathcal{B}(\mathcal{H}_1) \rightarrow \mathcal{B}(\mathcal{H}_2)$ is called $k$-positive if~$\mathrm{id}_{k} \otimes\Phi:M_k(\mathbb{C}) \otimes \mathcal{B}(H_1) \rightarrow M_k(\mathbb{C}) \otimes \mathcal{B}(H_2)$ is positive. If $\Phi$ is $k$-positive for all $k \in \mathbb{N}$, then $\Phi$ is called completely positive.
\end{dfnAlpha}
 By definition, every complete positive map is  positive. However, the converse is not always true. A standard example is the transposition map on matrices, which is positive but fails to be completely positive. The following theorem provides a necessary and sufficient condition for a map to be completely positive.

\begin{theoremAlpha}
    [Choi's Theorem \cite{book,article,CHOI1975285,kraus1983states,Sudarshan1985}]
A linear map $\Phi :\mathcal{B}(\mathcal{H}_1)\rightarrow \mathcal{B}(\mathcal{H}_2)$ is completely positive if and only if its Choi matrix $C_\Phi =\sum_{{i,j}=1}^{n} e_{ij} \otimes \Phi(e_{ij})$ is positive semidefinite.
\end{theoremAlpha}
\subsubsection{Complete Order Isomorphism}
A linear map $\Phi :M_n(\mathbb{C}) \to M_m(\mathbb{C})$ is called a complete order isomorphism if both $\Phi$ and $\Phi^{-1}$ are completely positive. An operator $X \in M_n(\mathbb{C}) \otimes M_n(\mathbb{C})$ is said to satisfy CP vs. positivity correspondence if the complete positivity of a linear map $\Phi :M_n(\mathbb{C}) \to M_m(\mathbb{C})$ is equivalent to the positivity of $(\rm{id} \otimes \Phi)X$ for all $m \in \mathbb{N}$~\cite{Kye}. 
\begin{theoremAlpha}[Kye, Paulsen~\cite{Kye,Paulsen}] \label{thm:Kye-Paulsen}
    Let $\sigma :M_n(\mathbb{C}) \to M_n(\mathbb{C})$ be a linear map. Then the Choi matrix $C_{\sigma}$ satisfies the CP vs. positivity correspondence if and only if $\sigma$ is a complete order isomorphism.
\end{theoremAlpha}
\subsection{Space of Super-maps}\label{s3}
Let $\mathcal{H}_i$ be finite dimensional Hilbert spaces for $i \!\in \!\{1,2,3,4\}$, and let $\C{B}(\mathcal{H}_i)$ denote the algebra of bounded linear operators on $\mathcal{H}_i$. We denote by \[L(\C{B}(\mathcal{H}_i), \C{B}(\mathcal{H}_j))\] the space of linear operators from $\C{B}(\mathcal{H}_i)$ to $\C{B}(\mathcal{H}_j)$, and by \[L[L(\C{B}(\mathcal{H}_1),\C{B}(\mathcal{H}_2)),L(\C{B}(\mathcal{H}_3),\C{B}(\mathcal{H}_4))]\] we denote the space of linear operators from $L(\C{B}(\mathcal{H}_1),\C{B}(\mathcal{H}_2))$ to $L(\C{B}(\mathcal{H}_3),\C{B}(\mathcal{H}_4))$. A linear map  $\Theta :L(\C{B}(\mathcal{H}_1),\C{B}(\mathcal{H}_2)) \rightarrow L(\C{B}(\mathcal{H}_3),\C{B}(\mathcal{H}_4))$ is called a super-map.
\\

For $n\in\{1,2,3,4\}$, the standard basis of the space $\C{B}(\mathcal{H}_{n})$ is denoted by $e^{\mathcal{H}_n}_{ij}$. Using the bra-ket notation, we have $e^{\mathcal{H}_n}_{ij}=\ketbra{i_{\mathcal{H}_n}}{j_{\mathcal{H}_n}}$, where $\{\ket{i_{\mathcal{H}_n}}\}$ is an orthonormal basis for $\mathcal{H}_n$. An orthonormal basis for $L(\C{B}(\mathcal{H}_1), \C{B}(\mathcal{H}_2))$ is the following\cite{Gour,SohailAHP2025}
\begin{align}
    \mathcal{E}^{\mathcal{H}_1 \rightarrow \mathcal{H}_2}_{ijkl}(A)
    &= \tr({e_{ij}^{\mathcal{H}_1}}^{\dagger}A)e^{\mathcal{H}_2}_{kl}, \label{canonical_basis_super-map}
\end{align}
where ``$\dagger$" denotes the hermitian conjugate.
\\

Let us denote by $\mathbb{CP}[1 \rightarrow 2]$ and $\mathbb{CP}[3 \rightarrow 4]$ the convex cone of completely positive maps of $L(\C{B}(H_1),\C{B}(H_2))$ and $L(\C{B}(H_3),\C{B}(H_4))$ respectively. 
\begin{definition*} 
    A super-map $\Theta :L(\C{B}(\C{H}_1),\C{B}(\C{H}_2)) \rightarrow L(\C{B}(\C{H}_3),\C{B}(\C{H}_4))$ is called CP-preserving if it maps CP maps in $\mathbb{CP}[1 \rightarrow 2]$ to the CP maps in $\mathbb{CP}[3 \rightarrow 4]$, i.e., $\Theta(\mathbb{CP}[1 \rightarrow 2]) \subseteq \mathbb{CP}[3 \rightarrow 4]$.
\end{definition*}
\begin{definition*} 
    A super-map $\Theta :L(\B(\C{H}_1),\B(\C{H}_2)) \rightarrow L(\B(\C{H}_3),\B(\C{H}_4))$ is called completely CP preserving if the map $\mathrm{id} \otimes \Theta : L(\B(K_1),\B(K_2)) \otimes L(\B(\C{H}_1),\B(\C{H}_2)) \rightarrow L(\B(K_1),\B(K_2)) \otimes L(\B(\C{H}_3),\B(\C{H}_4))$ is CP-preserving for for all choices of Hilbert spaces $K_1$ and $K_2$.
\end{definition*}

The spaces $L\left[L(\B(\mathcal{H}_1),\B(\mathcal{H}_2)), L(\B(\mathcal{H}_3),\B(\mathcal{H}_4))\right]$ and $L(\B(\mathcal{H}_1),\B(\mathcal{H}_2)) \otimes L(\B(\mathcal{H}_3),\B(\mathcal{H}_4))$ are isomorphic as vector spaces under the identification
\begin{eqnarray}
    \Theta \leftrightarrow \Lambda_{\Theta}:=\sum_{ijkl}  \mathcal{E}^{\mathcal{H}_1 \rightarrow \mathcal{H}_2}_{ijkl} \otimes \Theta( \mathcal{E}^{\mathcal{H}_1 \rightarrow \mathcal{H}_2}_{ijkl}).
\end{eqnarray}
It should be noted that the object $\Lambda_{\Theta}$ has the mathematical structure similar to Choi matrices.
\begin{theorem*}
    [\cite{Chiribella,Gour,SohailAHP2025}] The super-map $$\Theta :L(B(H_1),B(H_2)) \rightarrow L(B(H_3),B(H_4))$$ is completely CP-preserving if and only if $$\Lambda_{\Theta}:=\sum_{ijkl}  \mathcal{E}^{H_1 \rightarrow H_2}_{ijkl} \otimes \Theta( \mathcal{E}^{H_1 \rightarrow H_2}_{ijkl}) $$ is CP. \label{Choi-type_theorem}
\end{theorem*}

%We will refer to $\Lambda_{\Theta}$ as the Choi-type representation of the super-map $\Theta$ in this paper. 
\section{Main Results} \label{sec:Main Results}
 In this section, we present the formal statements and proofs of the main results outlined in the introduction. We begin with the structural properties of the convolution algebra on $\mathbb{C}_0[S]$ and proceed to the Fourier inversion and Plancherel formulas for finite inverse semigroups, followed by our Bochner-type theorem and its reduction to the Choi's theorem on complete positivity.
\\

 Let $G$ be a finite group with cardinality $|G|$, and  $\mathbb{C}[G]$ denotes the group algebra of $G$ over the complex field $\mathbb{C}$.  A generic element $a \in \mathbb{C}[G]$ has the form $a= \sum_{i=1} ^{|G|} \alpha_i g_i$. So, under the identification $a \leftrightarrow ( \alpha_1, \alpha_2,....,\alpha_{|G|} ) $, the two spaces $\mathbb{C}[G]$ and $L(\mathbb{C}_0[G],\mathbb{C})$ are isomorphic as vector spaces. Now there is a natural multiplication defined on $\mathbb{C}[G]$, induced by the group multiplication, namely
  \begin{align}
      a_1\cdot a_2 := \sum_{i,j=1}^{|G|} f_1(g_i)f_2(g_j) g_i g_j.
  \end{align}
  Using the fact that in group multiplication table, each element of $G$ appears exactly once in every row and every column, the above sum can be rewritten as 
  \begin{align}
      a_1\cdot a_2 &:= \sum_{k=1}^{|G|} \sum_{i=1}^{|G|} f_1(g_i)f_2(g_i ^{-1} g_k) g_k \nonumber \\
      &=\sum_{k=1}^{|G|}  (f_1*f_2)( g_k) g_k,
  \end{align}
   where \[(f_1*f_2)(g_k):= \sum_{i=1}^{|G|} f_1(g_i)f_2(g_i ^{-1} g_k)\] is  the convolution of $f_1$ and $f_2$. 
  %It can easily be verified that the convolution makes $L(G,C)$ an algebra. Hence 
  This shows that the convolution algebra $L(\mathbb{C}_0[G],\mathbb{C})$ is isomorphic to the group algebra $\mathbb{C}_0[G]$. This observation admits a natural generalization as follows:
\begin{theoremAlpha} \label{group_convolution}
    Let $G$ be a finite group with cardinality $d:=|G|$, and $\mathbb{C}[G]$ denote its group algebra. Let $\widetilde{\mathcal{A}}$ be an arbitrary algebra. Then the convolution algebra $L(\mathbb{C}[G],\widetilde{\mathcal{A}})$ of linear maps from $\mathbb{C}[G]$ to $\widetilde{\mathcal{A}}$ is isomorphic to the tensor product algebra $\mathbb{C}[G] \otimes \widetilde{\mathcal{A}}$, with the convolution defined as $\mathrm{\Phi} * \mathrm{\Phi}'(g_k)=\sum_i \mathrm{\Phi}(g_i) \mathrm{\Phi}'(g_i ^{-1} g_k)$.
\end{theoremAlpha}
See Appendix~\ref{Appendix:Proof of Theorem A} for the proof.
Now we generalize the above theorem in the context of contracted semigroup algebra. We begin by defining convolution in the space $L(\mathbb{C}_0[S], \mathcal{A})$ of linear maps from $\mathbb{C}_0[S]$ to an arbitrary algebra $\mathcal{A}$.
\begin{dfn}
    We define the convolution of two maps $\mathrm{\Phi}$ and $\mathrm{\Phi'}$ in the space $L\big(\mathbb{C}_{0}[S], \mathcal{A}\big)$ as follows:
 \begin{align}
    \mathrm{\Phi} * \mathrm{\Phi'}(s_k) := 
       \begin{cases}
                \sum_{i,j}\mathrm{\Phi}(s_i) \mathrm{\Phi'}(s_j)   &  \text{if}\ s_k \in \mathrm{Im}(\cdot) , \\ 
                0 & \text{if}\ s_k \notin \mathrm{Im}(\cdot),
       \end{cases}\label{semigroup_convolution_dfn}
 \end{align}
 for all $s_k \in S \setminus \{z\}$, where the sum is over the pairs $(i,j)$ such that $s_i \cdot s_j=s_k$, and $\mathrm{Im}(\cdot)$ denotes the image of the multiplication map $\cdot : S \times S \rightarrow S$ of the semigroup $S$.
\end{dfn}
\begin{theorem} \label{thm:semigroup convolution theorem}
    Let $S$ be a semigroup with zero element $z$ and denote the contracted algebra of $S$ by $\mathbb{C}_{0}[S]$. Then the convolution algebra $L(\mathbb{C}_{0}[S], \mathcal{A})$ is isomorphic to $\mathbb{C}_{0}[S] \otimes \mathcal{A}$, where the convolution product is defined by Eq.~(\ref{semigroup_convolution_dfn}).
\end{theorem}
\begin{prof}
Let us consider a linear map $\mathrm{\Phi}:\mathbb{C}_{0}[S] \rightarrow \mathcal{A}$, where $\mathcal{A}$ is an arbitrary algebra. We denote the space of all such maps by $L\big(\mathbb{C}_{0}[S], \mathcal{A}\big)$. Again, it can be shown that the spaces $L\big(\mathbb{C}_{0}[S], \mathcal{A}\big)$ and $\mathbb{C}_{0}[S] \otimes \mathcal{A}$ are isomorphic under the following identification:
\begin{align}
    \mathrm{\Phi} \leftrightarrow \sum_{s_i \neq 0} s_{i} \otimes \mathrm{\Phi}(s_{i}). \label{semigroup_CJ}
\end{align}
With the convolution product defined in Eq.~(\ref{semigroup_convolution_dfn}), the space $L\big(\mathbb{C}_{0}[S], \mathcal{A}\big)$ is a convolution algebra. We have the tensor product algebra $\mathbb{C}_{0}[S] \otimes \mathcal{A}$ where the multiplication rule comes from the individual algebras $\mathbb{C}_{0}[S]$ and $\mathcal{A}$. We now show that the identification in Eq.~(\ref{semigroup_CJ}) is an algebra isomorphism. Let $x$ and $x'$ be two arbitrary elements in $\mathbb{C}_{0}[S] \otimes \mathcal{A}$ and let $\mathrm{\Phi}$ and $\mathrm{\Phi}'$ be the corresponding maps in $L(\mathbb{C}_{0}[S],\mathcal{A})$. Now we have
 \begin{align}
     \nonumber
         x \cdot x' &= \bigg(\sum_{s_i \neq 0} s_i \otimes \mathrm{\Phi}(s_i)\bigg) \cdot \bigg( \sum_{s_j \neq 0} s_j \otimes \mathrm{\Phi}'(s_j)\bigg) \\
         &= \sum_{\substack{s_i \neq 0 \\
         s_j \neq 0}} s_{i}s_{j} \otimes \mathrm{\Phi}(s_i) \mathrm{\Phi}'(s_j). \label{semigroup_table}
     \end{align}
     Now if we look at the term $s_{i}s_{j}$, it can either be $ s_{k} \in \mathrm{Im}(\cdot)$ for some $k$ or it can be $0$. In the above summation, we collect all those terms for which $s_{i}s_{j}=s_{k} \neq 0$ and rewrite Eq.~(\ref{semigroup_table}) as follows:
     \begin{align}
         x \cdot x' &= \sum_{\substack{s_k \neq 0\\s_{k} \in \mathrm{Im}(\cdot)}} s_{k} \otimes \left( \sum_{p,q}\mathrm{\Phi}(s_p)\mathrm{\Phi}'(s_q) \right)+ 0 \otimes \bigg( \sum_{u,v}\mathrm{\Phi}(s_u)\mathrm{\Phi}'(s_v) \bigg), \label{semigroup_table_1}
     \end{align}
    where the summation inside the bracket of the first term is over those pair of elements $(s_{p}, s_{q})$ such that $s_{p} s_{q}= s_{k} \neq 0$ and the summation inside the bracket of the second term is over the pairs $(u,v)$ such that $s_u \cdot s_v= 0$. As the second term vanishes, we have
    \begin{align}
         x \cdot x' &= \sum_{\substack{s_k \neq 0\\s_{k} \in \mathrm{Im}(\cdot)}} s_{k} \otimes \left( \sum_{p,q}\mathrm{\Phi}(s_p)\mathrm{\Phi}'(s_q) \right) \nonumber\\
         &= \sum_{\substack{s_k \neq 0\\s_{k} \in \mathrm{Im}(\cdot)}} s_{k} \otimes \left( \sum_{p,q}\mathrm{\Phi}(s_p)\mathrm{\Phi}'(s_q) \right)+ \sum_{\substack{s_k \neq 0\\s_{k} \notin \mathrm{Im}(\cdot)}} s_{k} \otimes 0 \label{semigroup_table_3} \\
         &= \sum_{\substack{s_k \neq 0\\s_{k} \in S}} s_{k} \otimes (\mathrm{\Phi} * \mathrm{\Phi'})(s_{k}),\label{semigroup_table_2}
     \end{align}
    where we have used the definition of convolution~(\ref{semigroup_convolution_dfn}) in the last line. This completes the proof.
\end{prof}
\\

Let us now consider a $d$-dimensional Hilbert space $\mathcal{H}$, and denote the matrix units of $\mathcal{B(H)}$ by $\{e_{ij}\}_{i,j=1}^d$. The set $S:=\{e_{ij}\}_{i,j=1}^d \cup \{ 0\} \subset \mathcal{B}(\mathcal{H})$ with ``$0$" being the zero matrix, is a semigroup under the matrix multiplication:
\begin{align}
    e_{ij} \cdot e_{kl}= e_{il} \delta_{jk}. \label{matrix_unit_mult}
\end{align}
 Then the contracted algebra $\mathbb{C}_{0}[S]$ is isomorphic to the full algebra $\mathcal{B}(\mathcal{H})$ \cite{Clifford_semigroup, Amitsur_1951}.
Hence, we can view the matrix algebra $\mathcal{B}(\mathcal{H})$ as the contracted semigroup algebra of the semigroup $\{e_{ij}\}_{i,j=1}^d \cup \{ 0\}$. In the specific case where we consider the space $L(\mathcal{B}(\mathcal{H}_1), \mathcal{B}(\mathcal{H}_2))$ of linear maps from $\mathcal{B}(\mathcal{H}_1)$ to $\mathcal{B}(\mathcal{H}_2)$, Theorem~\ref{thm:semigroup convolution theorem} becomes the following:
\\
The convolution algebra $L(\mathcal{B}(\mathcal{H}_1), \mathcal{B}(\mathcal{H}_2))$ and the algebra $\mathcal{B}(\mathcal{H}_1) \otimes \mathcal{B}(\mathcal{H}_2)$ are isomorphic, with the isomorphism being the Choi-Jamiołkowski isomorphism $$\Phi \mapsto \sum_{i,j} e_{ij} \otimes \Phi(e_{ij}),$$ and the convolution product being 
\begin{align}
    \Phi * \Phi' (e_{ij}):= \sum_{k} \Phi(e_{ik}) \Phi'(e_{kj}), \label{eq:convolution dfn matrix algebra}
\end{align}
 and thereby retrieving the result of \cite{Sohail_2022}.
\\

The basis $ \mathcal{E}^{\mathcal{H}_{1} \rightarrow \mathcal{H}_{2}}_{ijkl}$ for the space $L(\mathcal{B}(\mathcal{H}_1), \mathcal{B}(\mathcal{H}_2))$ is defined in Eq.~(\ref{canonical_basis_super-map}). Now we show that $\Big\{ \mathcal{E}^{\mathcal{H}_{1} \rightarrow \mathcal{H}_{2}}_{ijkl}\Big\} \cup \big\{0\big\}$ forms a semigroup.
\begin{proposition}\label{prop:semigroup of basis map}
     The subset  $S:=\Big\{ \mathcal{E}^{\mathcal{H}_{1} \rightarrow \mathcal{H}_{2}}_{ijkl}\Big\} \cup \big\{0\big\} \subset L(\mathcal{B}(\mathcal{H}_1), \mathcal{B}(\mathcal{H}_2))$ forms a semigroup under the convolution product defined in Eq.~(\ref{eq:convolution dfn matrix algebra}).
\end{proposition}
\begin{prof}
The convolution between two elements $\mathcal{E}^{\mathcal{H}_{1} \rightarrow \mathcal{H}_{2}}_{ijkl} $ and $ \mathcal{E}^{\mathcal{H}_{1} \rightarrow \mathcal{H}_{2}}_{pqrs}$ is given by
\begin{align}
\nonumber
     \mathcal{E}^{\mathcal{H}_{1} \rightarrow \mathcal{H}_{2}}_{ijkl} * \mathcal{E}^{\mathcal{H}_{1} \rightarrow \mathcal{H}_{2}}_{pqrs} (e^{\mathcal{H}_{1}}_{uv}) &=\sum_{w}  \mathcal{E}^{\mathcal{H}_{1} \rightarrow \mathcal{H}_{2}}_{ijkl}(e^{\mathcal{H}_{1}}_{uw})  \mathcal{E}^{\mathcal{H}_{1} \rightarrow \mathcal{H}_{2}}_{pqrs} (e^{\mathcal{H}_{1}}_{wv}) \\
\nonumber     
     &=\sum_{w} \tr(e^{\mathcal{H}_{1} \dagger}_{kl} e^{\mathcal{H}_{1}}_{uw}) \tr(e^{\mathcal{H}_{1} \dagger}_{rs} e^{\mathcal{H}_{1}}_{wv}) e^{\mathcal{H}_{2}}_{ij} e^{\mathcal{H}_{2}}_{pq}\\
\nonumber      
     &=\sum_{w} \delta_{ku} \delta_{lw} \delta_{rw} \delta_{sv} \delta_{jp} e^{\mathcal{H}_{2}}_{iq}\\
\nonumber      
     &=\delta_{ku} \delta_{sv} \delta_{lr}\delta_{jp} e^{\mathcal{H}_{2}}_{iq}\\
\nonumber      
     &=\tr(e^{\mathcal{H}_{1} \dagger}_{ks} e^{\mathcal{H}_{1}}_{uv}) e^{\mathcal{H}_{2}}_{iq} \delta_{lr}\delta_{jp}\\
\nonumber      
     &=\mathcal{E}^{\mathcal{H}_{1} \rightarrow \mathcal{H}_{2}}_{iqks} (e^{\mathcal{H}_{1}}_{uv})\delta_{lr}\delta_{jp}
\end{align} 
which implies that $\mathcal{E}^{\mathcal{H}_{1} \rightarrow \mathcal{H}_{2}}_{ijkl} * \mathcal{E}^{\mathcal{H}_{1} \rightarrow \mathcal{H}_{2}}_{pqrs} =\mathcal{E}^{\mathcal{H}_{1} \rightarrow \mathcal{H}_{2}}_{iqks} \delta_{lr}\delta_{jp}$. This completes the proof.
\end{prof}
\\

Recalling that $L(\mathcal{B}(\mathcal{H}_1),\mathcal{B}(\mathcal{H}_2))$ and $L(\mathcal{B}(\mathcal{H}_3),\mathcal{B}(\mathcal{H}_4))$ are convolution algebras with the convolution being defined as in Eq.~(\ref{eq:convolution dfn matrix algebra}), and taking in to account Proposition~\ref{prop:semigroup of basis map}, we define convolution on $\scalemath{0.95}{L\left[L(\mathcal{B}(\mathcal{H}_1),\mathcal{B}(\mathcal{H}_2)), L(\mathcal{B}(\mathcal{H}_3),\mathcal{B}(\mathcal{H}_4))\right]}$ as follows:
\begin{align} \label{eq:convolution dfn for supermaps}
    \Theta_1 \star \Theta_2 (\mathcal{E}^{\mathcal{H}_{1} \rightarrow \mathcal{H}_{2}}_{ijkl}):= \sum_{pq} \Theta_1(\mathcal{E}^{\mathcal{H}_{1} \rightarrow \mathcal{H}_{2}}_{ipkq}) * \Theta_2(\mathcal{E}^{\mathcal{H}_{1} \rightarrow \mathcal{H}_{2}}_{pjql}).
\end{align}
In this setting Theorem~\ref{thm:semigroup convolution theorem} becomes the following:
\\
The convolution algebra $\scalemath{0.95}{L\left[L(\mathcal{B}(\mathcal{H}_1),\mathcal{B}(\mathcal{H}_2)), L(\mathcal{B}(\mathcal{H}_3),\mathcal{B}(\mathcal{H}_4))\right]}$ with the convolution being defined by Eq.~(\ref{eq:convolution dfn for supermaps}) and the convolution algebra $L(\mathcal{B}(\mathcal{H}_1),\mathcal{B}(\mathcal{H}_2)) \otimes L(\mathcal{B}(\mathcal{H}_3),\mathcal{B}(\mathcal{H}_4))$ with the convolution defined by Eq.~(\ref{eq:convolution dfn matrix algebra}) are isomorphic under the identification $\Theta \mapsto \sum_{ijkl}  \mathcal{E}^{H_1 \rightarrow H_2}_{ijkl} \otimes \Theta( \mathcal{E}^{H_1 \rightarrow H_2}_{ijkl})$.

 A super-map $$\mathrm{\Theta} : L\left(\mathcal{B}\left(\mathcal{H}_{1}\right),\mathcal{B}\left(\mathcal{H}_{2}\right)\right)\rightarrow L\left(\mathcal{B}\left(\mathcal{H}_{3}\right),\mathcal{B}\left(\mathcal{H}_{4}\right)\right)$$ induces a map $$T: \mathcal{B}\left(\mathcal{H}_{1}\right) \otimes \mathcal{B}\left(\mathcal{H}_{2}\right) \rightarrow \mathcal{B}\left(\mathcal{H}_{3}\right)\otimes \mathcal{B}\left(\mathcal{H}_{4}\right) $$ at the level of Choi matrices, known as the “representing map” \cite{Chiribella} of $\Theta$ and it is given by
\begin{equation}
            T(X)= C_{\mathrm{\Theta}\left(\Gamma_{X}\right)}, \label{equ:rep_map2}
\end{equation}
where $X \in \mathcal{B}\left(\mathcal{H}_{1}\right) \otimes \mathcal{B}\left(\mathcal{H}_{2}\right)$ is the Choi matrix of  $\Gamma_{X}$, and $C_{\mathrm{\Theta}\left(\Gamma_{X}\right)}$ is the Choi matrix for map $\mathrm{\Theta}\left(\Gamma_{X}\right)$. Now, under the identification: $\mathrm{\Theta} \leftrightarrow T$, we have~\cite{Chiribella, Gour, sohail_2023_June}
\begin{align*}
    L[L\left(\mathcal{B}\left(\mathcal{H}_{1}\right),\mathcal{B}\left(\mathcal{H}_{2}\right)\right),L\left(\mathcal{B}\left(\mathcal{H}_{3}\right),\mathcal{B}\left(\mathcal{H}_{4}\right)\right)] \cong L[\mathcal{B}\left(\mathcal{H}_{1}\right) \otimes \mathcal{B}\left(\mathcal{H}_{2}\right),\mathcal{B}\left(\mathcal{H}_{3}\right)\otimes \mathcal{B}\left(\mathcal{H}_{4}\right)].
\end{align*}
 We now prove that the map $\mathrm{\Theta} \leftrightarrow T$ preserves the respective convolutions. Let $T \leftrightarrow \Theta_1 \star \Theta_2$. Then we aim to show that $T= T_1 * T_2$:
\\
\begin{align}
    T\left(e^{\mathcal{H}_{1}}_{ij} \otimes e^{\mathcal{H}_{2}}_{kl}\right)&=T\left(C_{\mathcal{E}^{\mathcal{H}_{1} \rightarrow \mathcal{H}_{2}}_{klij}}\right) \nonumber \\
    &=C_{\Theta_1 \star \Theta_2\left({\mathcal{E}^{\mathcal{H}_{1} \rightarrow \mathcal{H}_{2}}_{klij}}\right)} \nonumber \\
    &= \sum_{pq} C_{\Theta_1\left({\mathcal{E}^{\mathcal{H}_{1} \rightarrow \mathcal{H}_{2}}_{kpiq}}\right)} C_{\Theta_2\left({\mathcal{E}^{\mathcal{H}_{1} \rightarrow \mathcal{H}_{2}}_{plqj}}\right)} \nonumber\\
    &= \sum_{pq} T_{1}\Big( e^{\mathcal{H}_{1}}_{iq} \otimes e^{\mathcal{H}_{2}}_{kp}\Big) T_{2}\Big( e^{\mathcal{H}_{1}}_{qj} \otimes e^{\mathcal{H}_{2}}_{pl}\Big) \nonumber \\
    &=T_1 *T_2 \Big( e^{\mathcal{H}_{1}}_{ij} \otimes e^{\mathcal{H}_{2}}_{kl}\Big).
\end{align} 
With the above observation, it follows that the convolution algebras $L[L(\mathcal{B}(\mathcal{H}_{1}),\mathcal{B}(\mathcal{H}_{2})),L(\mathcal{B}(\mathcal{H}_{3}),\mathcal{B}(\mathcal{H}_{4}))]$ and $L[\mathcal{B}\left(\mathcal{H}_{1}\right) \otimes \mathcal{B}\left(\mathcal{H}_{2}\right),\mathcal{B}\left(\mathcal{H}_{3}\right)\otimes \mathcal{B}\left(\mathcal{H}_{4}\right)]$ are isomorphic.
\subsection{Fourier Transform of Maps} \label{subsec:Fourier transform of maps}
Let $G$ be a finite group, and $f: G \rightarrow \mathbb{C}$ be a complex-valued function on $G$. Let $\rho: G \rightarrow M_{d_{\rho}}(\mathbb{C})$ be a representation of the group $G$ of degree $d_{\rho}$. The Fourier transform of the function $f$ with respect to the representation $\rho$ is given by:
\begin{align}
    \widehat{f}(\rho)= \sum_{g \in G} f(g) \rho(g).
\end{align}
The inverse Fourier transform is given by 
\begin{align}
    f(g)= \frac{1}{|G|} \sum_{\rho \in \rm{Irr}(G)}\!\!\!\! d_{\rho} \tr \big( \widehat{f}(\rho) \rho (g^{-1})\big).
\end{align}
If, instead of the complex valued function $f: G \rightarrow \mathbb{C}$, we consider a map $\Phi : G \rightarrow \mathcal{A}$, where $\mathcal{A}$ is an arbitrary algebra, it is natural to define the Fourier transform of $\Phi$ with respect to a representation $\rho: G \rightarrow M_{d_{\rho}}(\mathbb{C})$ as follows:
\begin{align}
    \widehat{\Phi}(\rho)=  \sum_{g \in G} \rho(g) \otimes \Phi(g) \in M_{d_{\rho}}(\mathbb{C}) \otimes \mathcal{A}.\label{F.T_group}
\end{align}
The inverse Fourier transform is given by
\begin{align} \label{eq:Fourier inversion map for group}
     \Phi(g)= \frac{1}{|G|} \sum_{\rho \in \mathrm{Irr}(G)} \!\!\!\!d_{\rho} \tr_{d_{\rho}} \big[(\rho (g^{-1}) \otimes \mathbb{I}) \widehat{\Phi}(\rho) \big],
\end{align}
where $\tr_{d_{\rho}}$ denotes partial trace over $M_{d_{\rho}}(\mathbb{C})$, and $\rm{Irr}(G)$ denotes a complete set of inequivalent irreducible representations of $G$. The Plancherel formula is given by (See Appendix~\ref{appendix:Proof of Plancherel formula for finite groups} for a proof):
\begin{align} \label{eq: Plancheral formula for finite group}
    \sum_{g \in G} \Phi(g^{-1}) \Psi(g) &=  \frac{1}{|G|} \sum_{\rho \in \rm{Irr}(G)} d_{\rho} \tr_{d_{\rho}} \big[ \widehat{\Phi}(\rho) \widehat{\Psi}(\rho)  \big].
\end{align}
\\

Using the definition of convolution used in Theorem~\ref{group_convolution}, one can prove that the Fourier transform of the convolution of two maps on a finite group is the product of their respective Fourier transform.
\subsubsection{Fourier Transform for Linear Maps on Contracted Semigroup Algebras}
 Let $S$ be a finite semigroup and $z \in S$ be the zero element. The Fourier transform of a linear map $\Phi: \mathbb{C}_{0}[S] \rightarrow \mathcal{A}$ with respect to a representation $\rho:\mathbb{C}_{0}[S] \rightarrow M_{d_{\rho}}(\mathbb{C})$ is defined as
\begin{align}
    \widehat{\Phi}(\rho)=\sum_{\substack{s \neq 0 \\ s \in S}} \rho(s) \otimes \Phi(s) \in M_{d_{\rho}}(\mathbb{C}) \otimes \mathcal{A}.
\end{align}

Now we prove the convolution theorem for the Fourier transform of maps on the contracted semigroup algebras.
\begin{proposition}
    Let $\mathbb{C}_0[S]$ be the contracted semigroup algebra of a semigroup $S$ and $\mathcal{A}$ be an arbitrary algebra. Then for maps $\Phi: \mathbb{C}_{0}[S] \rightarrow \mathcal{A}$ and $\Phi': \mathbb{C}_{0}[S] \rightarrow \mathcal{A}$, and a representation $\rho:\mathbb{C}_{0}[S] \rightarrow M_{d_{\rho}}(\mathbb{C})$,  we have 
    \begin{align}
    \widehat{\Phi * \Phi'}(\rho)= \widehat{\Phi}(\rho) \widehat{\Phi'}(\rho).\label{convolution_theorem_for_semigroup}
\end{align}
%where the product in the right-hand side of Eq.~(\ref{convolution_theorem_for_semigroup}) is the product in the tensor product algebra $M_n(\mathbb{C}) \otimes \mathcal{A}$.
\end{proposition}
\begin{prof}
   We begin with the product of the Fourier transforms $\widehat{\Phi}(\rho)$ and $ \widehat{\Phi'}(\rho)$, \emph{i.e.,}
   \begin{align}
       \widehat{\Phi}(\rho) \widehat{\Phi'}(\rho) &=\bigg(\sum_{s \in S, s \neq 0} \rho(s) \otimes \mathrm{\Phi}(s)\bigg) \cdot \bigg( \sum_{t \in S, t \neq 0} \rho(t) \otimes \mathrm{\Phi}'(t)\bigg) \nonumber\\
         &=(\rho \otimes \mathrm{id}) \sum_{\substack{s,t \in S\nonumber\\
         s,t \neq 0}} st \otimes \Phi(s) \Phi'(t) \nonumber\\
       &=(\rho \otimes \mathrm{id})  \sum_{\substack{u \neq 0\\u \in S}} u \otimes (\mathrm{\Phi} * \mathrm{\Phi'})(u) \nonumber\\
       &=\widehat{\Phi * \Phi'}(\rho), \nonumber
   \end{align}
   where the 2nd equality follows from the fact that $\rho$ is a representation, and the 3rd equality follows from the arguments used in Eqs.~(\ref{semigroup_table}), (\ref{semigroup_table_1}), (\ref{semigroup_table_3}) and (\ref{semigroup_table_2}).
   This completes the proof.
\end{prof}
\\

Now we may ask whether there exists a one-to-one correspondence between $\Phi$ and  $\widehat{\Phi}(\rho)$. In general, such a correspondence does not exist. However, if $S$ is a finite inverse semigroup, and one considers a complete set of irreducible representations $\rho$ of $\mathbb{C}_0[S]$, then Wedderburn-Artin theorem yields such a correspondence.

The inverse semigroup algebra $\mathbb{C}[S]$ is semisimple~\cite{Munn_1957}, and $\mathbb{C}_0[S]$ is semisimple if and only if $\mathbb{C}[S]$ is semisimple~\cite{Clifford_semigroup}. Consequently there is an algebra isomorphism $$\mathbb{C}_0[S] \cong \bigoplus_{\rho \in \mathrm{Irr}(\mathbb{C}_0[S])} M_{d_{\rho}}(\mathbb{C}),$$ where $\mathrm{Irr}(\mathbb{C}_0[S])$ denotes a complete set of inequivalent irreducible representations of the contracted algebra $\mathbb{C}_0[S]$. Identifying $\mathbb{C}_0[S]$ with the  convolution algebra $L(\mathbb{C}_0[S],\mathbb{C})$, one obtains
$$L(\mathbb{C}_0[S],\mathbb{C}) \cong \bigoplus_{\rho \in \mathrm{Irr}(\mathbb{C}_0[S])} M_{d_{\rho}}(\mathbb{C}),$$ with the explicit correspondence
\[\Phi \mapsto \bigoplus_{\rho \in \mathrm{Irr}(\mathbb{C}_0[S])} \sum_{s \in S \setminus \{0\}} \Phi(s) \rho(s) .\] This simple observation extends naturally to $\mathcal{A}$-valued functions, yielding
\begin{align}
    L(\mathbb{C}_0[S],\mathcal{A}) \cong \bigoplus_{\rho \in \mathrm{Irr}(\mathbb{C}_0[S])} M_{d_{\rho}}(\mathbb{C}) \otimes \mathcal{A}
\end{align}
 with the correspondence
 \begin{align}
     \Phi \mapsto \!\!\!\bigoplus_{\rho \in \mathrm{Irr}(\mathbb{C}_0[S])} \sum_{s \in S \setminus \{0\}}  \rho(s) \otimes \Phi(s).
 \end{align}
Equivalently, one can say that there is one-to-one correspondence between $\Phi$ and $\bigoplus_{\rho \in \mathrm{Irr}(\mathbb{C}_{0}[S])} \widehat{\Phi}(\rho)$.

\subsection{Fourier Inversion for Finite Inverse Semigroups} \label{subsec:Fourier Inversion for finite inverse semigroup}
In this section, we derive a Fourier inversion formula for finite inverse semigroup. The proof relies on the following idea: Steinberg's isomorphism provides a way to construct a complete set of inequivalent irreducible representations of the contracted algebra of a finite inverse semigroup in terms of complete sets of inequivalent irreducible representations of the maximal subgroups associated with the $\mathcal{D}$-classes of the inverse semigroup (see Theorem~\ref{lemma:Stineberg_1}). The Fourier inversion formula for these maximal subgroups then helps to provide the Fourier inversion for the inverse semigroup. While the Fourier inversion formula for complex valued functions on the semigroup algebra of a finite inverse semigroup was established in~\cite{Malandro}, we extend this approach to derive a Fourier inversion formula for matrix-valued functions on the contracted algebra of a finite inverse semigroup.
\begin{theorem}\label{thm:fourier_inversion_inverse_semigroup}
    Let $\mathbb{C}_0[S]$ be the contracted semigroup algebra of a finite inverse semigroup $S$. Let $\rm{Irr}(\mathbb{C}_0[S])$ denote a complete set of inequivalent irreducible representations of $\mathbb{C}_0[S]$. Let $\{D_k \}_{k=1}^n$ be the set of $\mathcal{D}$-classes of $S$, let $r_k$ be the number of idempotents in $D_k$, and let $G_{e_k}$ be the maximal subgroup at an arbitrarily chosen idempotent element $e_k \in D_k$. Then the inversion formula for the Fourier transform $$\widehat{\Phi}(\rho)=\sum_{s \in S, s\neq 0} \rho(s) \otimes \Phi(s)=\sum_{s \in S, s\neq 0} \rho(\floor{s}) \otimes \widetilde{\Phi}(\floor{s})$$ of a linear map $\Phi: \mathbb{C}_0[S] \to M_n(\mathbb{C})$ is given by
   \begin{align}
    \widetilde{\Phi}(\floor{s})&=\frac{1}{r_k|G_{e_k}|} \sum_{\sigma \in \rm{Irr}(\mathbb{C}_{0}[S])}\!\!\!\!\!\! d_{\sigma} \tr_{d_{\sigma}}  \big[(\sigma(\floor{s^{-1}}) \otimes \mathbb{I}) \widehat{\Phi}(\sigma)\big],\label{eq:Fourier_inversion_for_maps_3}
\end{align}
where $d_\sigma$ denote the dimension of the representation $\sigma$. The linear maps $\widetilde{\Phi}:\mathbb{C}_0[S] \to M_n(\mathbb{C})$ and $\Phi$ are related by $\Phi(s)=\sum_{\substack{t \in S \\ t \geq s}} \mu(s,t) \widetilde{\Phi}(\floor{t})$, and $\widetilde{\Phi} (\floor{s})=\sum_{\substack{t \in S \\ t \geq s}} \Phi (t)$, where $\mu$ denotes the Möbius function on the natural partial order of $S$.
\end{theorem} 
\begin{prof}
     Let $\overline{\mathrm{Irr}}(\mathbb{C}_{0}[S])$ denote the complete set of inequivalent irreducible representations of $\mathbb{C}_0[S]$ constructed as in Theorem~\ref{lemma:Stineberg_1}. More precisely, a representation $\overline{\rho}_k \in \overline{\mathrm{Irr}}(\mathbb{C}_{0}[S])$ is given by
 \begin{align} \label{eq:representation of C_{0}}
     \overline{\rho}_k(\floor{s})=\begin{cases}
                                   E_{\rm{ran}(s),\rm{dom}(s)} \otimes \rho_{k}(p^{-1}_{\rm{ran}(s)}\, s \, p_{\rm{dom}(s)}) & \text{when} \quad s \in D_k, \\
                                   0 & \text{otherwise}.
                                  \end{cases}
 \end{align}
 \\
 Notice that the element $\sum_{s \in S, s\neq 0} s \otimes \Phi(s) \in \mathbb{C}_0[S] \otimes M_n(\mathbb{C})$ can also be expressed in terms of the groupoid basis as $\sum_{s \in S, s\neq 0} \floor{s} \otimes \widetilde{\Phi}(\floor{s})$. The linear maps $\widetilde{\Phi}$ and $\Phi$ are related to each other by the following relations:
 \begin{align}
     \Phi(s)&=\sum_{\substack{t \in S \\ t \geq s}} \mu(s,t) \widetilde{\Phi}(\floor{t}), \label{eq:Phi_to_Phi_tilde}\\
    \widetilde{\Phi} (\floor{s})&=\sum_{\substack{t \in S \\ t \geq s}} \Phi (t).\label{eq:Phi_tilde_to_Phi}
 \end{align}
 Since the Fourier transform of $\Phi$ can be written as $\widehat{\Phi}(\rho)= (\rho \otimes \mathrm{id}) \sum_{s \in S, s\neq 0} s \otimes \Phi(s)$, it follows that 
 \begin{align}
     \widehat{\Phi}(\rho)=\sum_{s \in S, s\neq 0} \rho(s) \otimes \Phi(s)=\sum_{s \in S, s\neq 0} \rho(\floor{s}) \otimes \widetilde{\Phi}(\floor{s}). \label{eq:Fourier transform wrt induced representation_0}
 \end{align}
 Now, for $\overline{\rho}_k \in \overline{\mathrm{Irr}}(\mathbb{C}_{0}[S])$,
 \begin{align}
     \widehat{\Phi}(\overline{\rho}_k)=\sum_{s \in D_k} E_{\rm{ran}(s),\rm{dom}(s)} \otimes \rho_{k}(p^{-1}_{\rm{ran}(s)}sp_{\rm{dom}(s)}) \otimes \widetilde{\Phi}(\floor{s}). \label{eq:Fourier transform wrt induced representation}
 \end{align}
 Observe that the Fourier transform $\widehat{\Phi}(\bar{\rho}_k)$ is a matrix indexed by the idempotents of $D_k$. For idempotents $a,b \in D_k$, the $(a,b)$-th entry of $\widehat{\Phi}(\bar{\rho}_k)$ is given by
 \begin{align}
    \left( \widehat{\Phi}(\bar{\rho}_k)\right)_{a,b}&= \sum_{\substack{s \in D_k \\\rm{ran}(s)=a\\ \rm{dom}(s)=b }} \rho_{k}\left(p^{-1}_{\rm{ran}(s)}s~p_{\rm{dom}(s)}\right) \otimes \widetilde{\Phi}(\floor{s}). \label{eq:Matrix elements of Fourier transform}
 \end{align}
 \\
 By Proposition~\ref{prop:D_k to G_(e_k)}, the element $u:=p^{-1}_{\rm{ran}(s)}s~p_{\rm{dom}(s)}$ is an element of the maximal subgroup $G_{e_k}$ at an idempotent $e_k \in D_k$. We can invert the relation $u=p^{-1}_{\rm{ran}(s)}s~p_{\rm{dom}(s)}$ as follows: $u=p^{-1}_{\rm{ran}(s)}s~p_{\rm{dom}(s)}
     \iff  p_{\rm{ran}(s)}u p^{-1}_{\rm{dom}(s)} =p_{\rm{ran}(s)}p^{-1}_{\rm{ran}(s)}sp_{\rm{dom}(s)}p^{-1}_{\rm{dom}(s)}
     \iff  p_{\rm{ran}(s)}u p^{-1}_{\rm{dom}(s)} =\rm{ran}(s) \, s \,\rm{dom}(s)
     =ss^{-1} s s^{-1}s
     =s$.
 Since $\rm{ran}(s)=a$ and $ \rm{dom}(s)=b$, we have, $s=p_a \, u \, p^{-1}_b$. With this observation, we can write the summation in Eq.~(\ref{eq:Matrix elements of Fourier transform}) as sum over those elements $u \in G_{e_k}$ such that $\rm{ran}(p_a \, u \, p^{-1}_b)=a$ and $\rm{dom}(p_a \, u \, p^{-1}_b)=b$:
 \begin{align}
    \left( \widehat{\Phi}(\bar{\rho}_k)\right)_{a,b}&=\!\!\!\!\!\!\!\!\!\!\!\! \sum_{\substack{u \in G_{e_k} \\ \rm{ran}(p_a \, u \, p^{-1}_b)=a,\\ \rm{dom}(p_a \, u \, p^{-1}_b)=b }}\!\!\!\!\!\!\!\!\!\!\!\! \rho_{k}(u) \otimes \widetilde{\Phi}(\floor{p_a \, u \, p^{-1}_b}). \label{eq:Matrix elements of Fourier transform_1}
 \end{align}
 \\
 By Proposition~\ref{prop:Transformation D_k to G_{e_k}}, we can rewrite Eq.~(\ref{eq:Matrix elements of Fourier transform_1}) as follows:
 \begin{align}
    \left( \widehat{\Phi}(\bar{\rho}_k)\right)_{a,b}&= \sum_{\substack{u \in G_{e_k} }} \rho_{k}(u) \otimes \widetilde{\Phi}(\floor{p_a \, u \, p^{-1}_b}). \label{eq:Matrix elements of Fourier transform_3}
 \end{align}
 We can immediately identify $ \left( \widehat{\Phi}(\bar{\rho}_k)\right)_{a,b}$ in Eq.~(\ref{eq:Matrix elements of Fourier transform_3}) as the Fourier transform of the map $\widetilde{\Phi}(\floor{p_a \, (\cdot) \, p^{-1}_b})$ on the maximal subgroup $G_{e_k}$. Hence, we use Eq.~(\ref{eq:Fourier inversion map for group}) for the Fourier inversion and arrive at the following:
 \begin{align}
      \widetilde{\Phi}(\floor{s})=\widetilde{\Phi}(\floor{p_a \, u \, p^{-1}_b})&=\frac{1}{|G_{e_k}|} \sum_{\rho_k \in \rm{Irr(G)}} d_{\rho_k} \tr_{d_{\rho_k}} \left[(\rho_k (u^{-1}) \otimes \mathbb{I}) \left( \widehat{\Phi}(\bar{\rho}_k)\right)_{a,b} \right], \label{eq:Fourier_inversion_for_maps}
 \end{align}
 where $a=\rm{ran}(s)$ and $b=\rm{dom}(s)$. It follows from Eq.~(\ref{eq:Fourier transform wrt induced representation}) that 
 \begin{align}
      \widehat{\Phi}(\bar{\rho}_k)=\sum_{i,j \in D_k} E_{i,j} \otimes \left( \widehat{\Phi}(\bar{\rho}_k)\right)_{i,j}, \label{eq:Fourier transform wrt induced representation_1}
 \end{align}
 where $i,j \in D_k$ are idempotents. A quick calculation shows 
 \begin{align}
     \tr_{d_{\rho_k}} \left[(\rho_k (u^{-1}) \otimes \mathbb{I}) \left( \widehat{\Phi}(\bar{\rho}_k)\right)_{a,b} \right] &= (\tr_{r_k} \otimes \tr_{d_{\rho_k}}) \big[(E_{b,a} \otimes\rho_k (u^{-1}) \otimes \mathbb{I}) \widehat{\Phi}(\bar{\rho}_k) \big]. \label{eq:quick_calculation}
 \end{align}
Since $s \in D_k$, from Eq.~(\ref{eq:representation of C_{0}}), we have
 \begin{align} 
     \overline{\rho}_k(\floor{s})&= E_{\rm{ran}(s),\rm{dom}(s)} \otimes \rho_{k}(p^{-1}_{\rm{ran}(s)}\, s \, p_{\rm{dom}(s)}).
 \end{align}
 Notice that $\rm{ran}(s)=\rm{dom}(s^{-1})$ and $\rm{dom}(s)=\rm{ran}(s^{-1})$. Recalling that $u^{-1}=p^{-1}_{\rm{dom}(s)}\,s^{-1} \,p_{\rm{ran}(s)}$, we have 
 \begin{align}
     \overline{\rho}_k(\floor{s^{-1}})&= E_{\rm{ran}(s^{-1}),\rm{dom}(s^{-1})} \otimes \rho_{k}(p^{-1}_{\rm{ran}(s^{-1})}\, s^{-1} \, p_{\rm{dom}(s^{-1})})\nonumber\\
     &= E_{\rm{dom}(s),\rm{ran}(s)} \otimes \rho_{k}(p^{-1}_{\rm{dom}(s)}\, s^{-1} \, p_{\rm{ran}(s)}) \nonumber\\
     &=E_{b,a} \otimes \rho_{k}(u^{-1}). \label{eq:Induced representation action on inverse groupoid}
 \end{align}
 With the above observation, Eq.~(\ref{eq:quick_calculation}) becomes the following:
 \begin{align}
     \tr_{d_{\rho_k}} \left[(\rho_k (u^{-1}) \otimes \mathbb{I}) \left( \widehat{\Phi}(\bar{\rho}_k)\right)_{a,b} \right] &= (\tr_{r_k} \otimes \tr_{d_{\rho_k}}) \big[(\overline{\rho}_k(\floor{s^{-1}}) \otimes \mathbb{I}) \widehat{\Phi}(\bar{\rho}_k) \big]. \label{eq:quick_calculation_1}
 \end{align}
 Notice that $\tr_{r_k} \otimes \tr_{d_{\rho_k}}= \tr_{d_{\overline{\rho}_k}}$. Using Eq.~(\ref{eq:quick_calculation_1}) in Eq.~(\ref{eq:Fourier_inversion_for_maps}), we have the following. 
 \begin{align}
      \widetilde{\Phi}(\floor{s})&=\frac{1}{|G_{e_k}|} \sum_{\rho_k \in \rm{Irr(G)}} d_{\rho_k} \tr_{d_{\overline{\rho}_k}}  \big[(\overline{\rho}_k(\floor{s^{-1}}) \otimes \mathbb{I}) \widehat{\Phi}(\bar{\rho}_k) \big].\label{eq:Fourier_inversion_for_maps_1}
 \end{align}
Let $\rm{Irr}(\mathbb{C}_{0}[S])$ be an arbitrary complete set of inequivalent irreducible representations of $\mathbb{C}_{0}[S]$. Let $\sigma \in \rm{Irr}(\mathbb{C}_{0}[S])$. Then $\sigma$ must be equivalent to an element $\overline{\rho}_k \in \overline{\mathrm{Irr}}(\mathbb{C}_{0}[S])$, i.e., there exists an invertible transformation $H$ such that $\overline{\rho}_k(s)= H^{-1} \sigma(s) H$ for all $s \in S$. From Eq.~(\ref{eq:Fourier transform wrt induced representation_0}), we have 
\begin{align}
     \widehat{\Phi}(\rho)&=\sum_{s \in S, s\neq 0} \overline{\rho}_{k}(\floor{s}) \otimes \widetilde{\Phi}(\floor{s}) \nonumber\\
     &=\sum_{s \in S, s\neq 0} H^{-1}\sigma(\floor{s}) H\otimes \widetilde{\Phi}(\floor{s}) \nonumber\\
     &=(H^{-1} \otimes \mathbb{I})\widehat{\Phi}(\sigma) (H \otimes \mathbb{I}).\label{eq:Fourier transform wrt induced representation_a}
\end{align}
Using the above relation in Eq.~(\ref{eq:Fourier_inversion_for_maps_1}), we get
\begin{align}
      \widetilde{\Phi}(\floor{s})&=\frac{1}{|G_{e_k}|} \sum_{\sigma \in \rm{Irr}(\mathbb{C}_{0}[S])} d_{\rho_k} \tr_{d_{\overline{\rho}_k}}  \big[(H^{-1}\sigma(\floor{s^{-1}}) H \otimes \mathbb{I}) (H^{-1} \otimes \mathbb{I})\widehat{\Phi}(\sigma) (H \otimes \mathbb{I})\big] \nonumber\\
      &=\frac{1}{|G_{e_k}|} \sum_{\sigma \in \rm{Irr}(\mathbb{C}_{0}[S])} d_{\rho_k} \tr_{d_{\sigma}}  \big[(HH^{-1}\sigma(\floor{s^{-1}}) HH^{-1} \otimes \mathbb{I}) \widehat{\Phi}(\sigma)\big] \nonumber\\
      &=\frac{1}{|G_{e_k}|} \sum_{\sigma \in \rm{Irr}(\mathbb{C}_{0}[S])} d_{\rho_k} \tr_{d_{\sigma}}  \big[(\sigma(\floor{s^{-1}}) \otimes \mathbb{I}) \widehat{\Phi}(\sigma)\big].\label{eq:Fourier_inversion_for_maps_2}
 \end{align}
Notice that by construction, the dimension of the induced representation $d_{\overline{\rho}_k}=r_k d_{\rho_k}$. Since $\overline{\rho}_k$ is equivalent to the representation $\sigma$, their dimensions must be the same, \emph{i.e.,} $d_{\sigma}=d_{\overline{\rho}_k}=r_k d_{\rho_k}$ and hence $d_{\rho_k}=\frac{d_{\sigma}}{r_k}$. With this observation, we obtain the following inversion formula
\begin{align}
    \widetilde{\Phi}(\floor{s})&=\frac{1}{r_k|G_{e_k}|} \sum_{\sigma \in \rm{Irr}(\mathbb{C}_{0}[S])} d_{\sigma} \tr_{d_{\sigma}}  \big[(\sigma(\floor{s^{-1}}) \otimes \mathbb{I}) \widehat{\Phi}(\sigma)\big].
\end{align}
This completes the proof.
\end{prof}
\\

Now we consider the finite dimensional matrix algebra $M_m(\mathbb{C})$. It can be verified that the set $S_{M}:=\{e_{ij}\}_{i,j=1}^m \cup \{0\}$ consisting of matrix units together with the zero matrix, forms an inverse semigroup. Moreover, the matrix algebra $M_m(\mathbb{C})$ is naturally identified with the contracted semigroup algebra of $S_M$, \emph{i.e.,} $\mathbb{C}_0[S_M]=M_m(\mathbb{C})$~\cite{Amitsur_1951}. The idempotent elements of $S_M$ are $e_{ii}$ for all $i \in \{1,2,...,m \}$ and $0$. The inverse of an element $e_{ij}$ is the $e_{ji}$ \emph{i.e.,} $e_{ij}^{-1}=e_{ji}$. It can also be verified that the maximal subgroup at an idempotent $e_{ii}$ is $G_{e_{ii}}=\{e_{ii}\}$.
\begin{propositionAlpha} \label{prop:1}
    The $\mathcal{D}$-classes of the inverse semigroup $S_M$ are $D_0:=\{ 0\}$ and $D_1:=\{ e_{ij} \mid 1\leq i,j \leq m\}$.
\end{propositionAlpha}
\begin{prof}
    Suppose that $e_{ij} \mathcal{D}~0$. Then there exist $e_{pq}$ such that $e_{qp}e_{pq}=0$ and $e_{pq}e_{qp}=e_{ij}e_{ji}$. However, for any matrix unit $e_{pq}$, one always has $e_{qp}e_{pq} \neq 0$, which yields a contradiction.  Hence, the $D$-class containing the zero matrix consists only of $0$. 
    
    Next, for arbitrary indices $i,j,k,l \in \{1,2,...,m \}$, consider the element $e_{ki}$. We have $e_{ki}^{-1}e_{ki}=e_{ij}e_{ij}^{-1}$ and $e_{ki}e_{ki}^{-1}=e_{kl}e_{kl}^{-1}$. Therefore, $e_{ij} \mathcal{D} e_{kl}$. This shows that all non-zero elements $e_{ij}$ lie in the same $\mathcal{D}$-class.
\end{prof}
\begin{corollary} \label{corollary:Fourier inversion becomes Choi inversion}
Let $S_{M}:=\{e_{ij}\}_{i,j=1}^m \cup \{0\}$ be the matrix unit inverse semigroup. Then the Fourier inversion formula in Eq.~(\ref{eq:Fourier_inversion_for_maps_3}) reduces to the Choi inversion formula in Eq.~(\ref{eq:Choi_inversion}).
\end{corollary}
\begin{prof}
    Recall that for two elements $s,t$ in an inverse semigroup $S$, one has $s\leq t$ if and only if there exists an idempotent $e \in S$ such that $s=et$ (see Definition~\ref{dfn:partil_ordering_in_inverse_semigroup}). Clearly, $0 \leq e_{ij}$ for all $i,j \in \{1,2,...,m \}$.
    
    Now, suppose that $e_{ij} \leq e_{kl}$. Then there exists an idempotent $e_{pp}$ such that $e_{ij}=e_{pp}e_{kl}=e_{pl} \delta_{pk}$. Since $e_{ij} \neq 0$, it follows that $p=k$, and hence $e_{ij}=e_{kl}$. Therefore the only element $x \in \{e_{ij}\}_{i,j=1}^m$ satisfying $e_{ij} \leq x$ is $x=e_{ij}$ itself.
    
    It follows from Eq.~(\ref{eq:groupoid_basis}) that the groupoid basis corresponding to $e_{ij}$ is $e_{ij}$ itself, \emph{i.e.,} $\floor{e_{ij}}=e_{ij}$. Consequently, by Eq.~(\ref{eq:Phi_tilde_to_Phi}), we have $\widetilde{\Phi}(\floor{e_{ij}})=\Phi(e_{ij})$. Since the $\mathcal{D}$-class $D_1$ contains $m$ idempotents, we have $r_1=m$. Moreover, $G_{e_{ii}}$ being a group with only one element in it, we have $|G_{e_{ii}}|=1$. 
    
    Substituting these observations into Eq.~(\ref{eq:Fourier_inversion_for_maps_3}), we obtain
    \begin{align}
    \widetilde{\Phi}(\floor{e_{ij}})&=\Phi(e_{ij})=\frac{1}{n} \sum_{\sigma \in \rm{Irr}(M_n(\mathbb{C}))} d_{\sigma} \tr_{d_{\sigma}}  \big[(\sigma(e_{ji}) \otimes \mathbb{I}) \widehat{\Phi}(\sigma)\big].
\end{align}
Now, up to equivalence, there is exactly one irreducible representation of the full matrix algebra $M_m(\mathbb{C})$~\cite{LangAlgebra}. We can ``take" this representation to be the identity map $``\rm{id}"$ on the matrix algebra $M_m(\mathbb{C})$ (under the identification of $M_m(\mathbb{C})$ with $\mathrm{End}(\mathbb{C}^m)$). Now using the fact that the dimension of this representation is $m$, \emph{i.e.,} $d_{\mathrm{id}}=m$ and the Fourier transform of the map $\Phi$ with respect to the representation $``\mathrm{id}"$ is the Choi matrix of $\Phi$, \emph{i.e.,} $\widehat{\Phi}(\mathrm{id})=C_{\Phi}$, we obtain 
\begin{align}
    \Phi(e_{ij})=\tr_m  \big[(e_{ji} \otimes \mathbb{I}) C_{\Phi}\big].
\end{align}
Equivalently, for an arbitrary $x \in M_{n}(\mathbb{C})$, we have
\begin{align}
    \Phi(x)=\tr_m  \big[(x^{\tau} \otimes \mathbb{I}) C_{\Phi}\big],
\end{align}
where $\tau$ denotes the transpose map. This completes the proof.
\end{prof}

\begin{theorem} \label{thm:Placheral formula for inverse semigroup}
    The Plancherel formula for a finite inverse semigroup $S$ is given by
    \begin{align}
        \sum_{\substack{s \in S\\s \neq 0}} r_{k} |G_{e_k}| \widetilde{\Phi}(\floor{s^{-1}}) \widetilde{\Psi}(\floor{s})&= \sum_{\sigma \in \rm{Irr}(\mathbb{C}_{0}[S])} d_{\sigma} \tr_{d_{\sigma}}[\widehat{\Phi}(\sigma)\widehat{\Psi}(\sigma)],
    \end{align}
    where $r_k$ and $G_{e_k}$ are defined in Theorem~\ref{thm:fourier_inversion_inverse_semigroup}.
\end{theorem}
\begin{prof}
    Since $s \in D_k$ if and only if $s^{-1} \in D_k$, the quantities $r_k$ and $|G_{e_k}|$ remain unchanged. Consequently, form Eq.~(\ref{eq:Fourier_inversion_for_maps_3}), we have 
    \begin{align*}
    \widetilde{\Phi}(\floor{s^{-1}})&=\frac{1}{r_k|G_{e_k}|} \sum_{\sigma \in \rm{Irr}(\mathbb{C}_{0}[S])} d_{\sigma} \tr_{d_{\sigma}}  \big[(\sigma(\floor{s}) \otimes \mathbb{I}) \widehat{\Phi}(\sigma)\big].
\end{align*}
 Now multiplying both the sides by $\widetilde{\Psi}(\floor{s})$ and summing over $s$ with $s \neq 0$, we obtain 
 \begin{align*}
     \sum_{\substack{s \in S\\ s \neq 0}} r_{k} |G_{e_k}| \widetilde{\Phi}(\floor{s^{-1}}) \widetilde{\Psi}(\floor{s})&= \sum_{\substack{s \in S\\ s \neq 0}}\sum_{\sigma \in \rm{Irr}(\mathbb{C}_{0}[S])} d_{\sigma} \tr_{d_{\sigma}}  \big[(\sigma(\floor{s}) \otimes \mathbb{I}) \widehat{\Phi}(\sigma)\big] \widetilde{\Psi}(\floor{s}) \\
     &=\sum_{\substack{s \in S\\ s \neq 0}} \sum_{\sigma \in \rm{Irr}(\mathbb{C}_{0}[S])} d_{\sigma} \tr_{d_{\sigma}}  \big[\widehat{\Phi}(\sigma)(\sigma(\floor{s}) \otimes \mathbb{I}) \big] \widetilde{\Psi}(\floor{s}) \\
     &=\sum_{\substack{s \in S\\ s \neq 0}} \sum_{\sigma \in \rm{Irr}(\mathbb{C}_{0}[S])} d_{\sigma} \tr_{d_{\sigma}}  \big[\widehat{\Phi}(\sigma)(\sigma(\floor{s}) \otimes \widetilde{\Psi}(\floor{s})) \big]  \\
     &= \sum_{\sigma \in \rm{Irr}(\mathbb{C}_{0}[S])} d_{\sigma} \tr_{d_{\sigma}}  \Big[\widehat{\Phi}(\sigma)\sum_{\substack{s \in S\\ s \neq 0}}(\sigma(\floor{s}) \otimes \widetilde{\Psi}(\floor{s})) \Big]  \\
     &= \sum_{\sigma \in \rm{Irr}(\mathbb{C}_{0}[S])} d_{\sigma} \tr_{d_{\sigma}}  \Big[\widehat{\Phi}(\sigma) \widehat{\Psi}(\sigma)\Big].  
\end{align*}
This completes the proof.
\end{prof}
\\
For matrix unit inverse semigroup $S_M$, we have $r_k=m$, $|G_{e_k}|=1$, $\widetilde{\Phi}(\floor{e_{ij}})=\Phi(e_{ij})$ and $\widetilde{\Psi}(\floor{e_{ij}})=\Psi(e_{ij})$. In this case, the Plancherel formula reduces to
\begin{align}
   \sum_{i,j=1}^m \Phi(e_{ji}) \Psi(e_{ij})& = \tr_{m}  \Big[\widehat{\Phi}(\rm{id}) \widehat{\Psi}(\rm{id})\Big]= \tr_m(C_{\Phi}C_{\Psi}).
\end{align}
\subsection{Positive Definite Maps and Bochner's theorem for Finite Inverse Semigroups} \label{subsec:Positive definite maps and Bochner's theorem for finite inverse semigroup}
Let $G$ be a finite group, and let $\mathcal{H}=\mathbb{C}^n$ be a Hilbert space. A map $\Phi:G \to \mathcal{B}(\mathcal{H})$ is called positive definite if the matrix $[\Phi(g^{-1}g')] \in \mathcal{B}(\mathcal{H}^{\bigoplus |G|})$ is positive semidefinite~\cite{Ozawa}, \emph{i.e.,}
\begin{align}
    \sum_{g,g' \in G} \inner{\Phi(g^{-1}g')h_{g'},h_g} \geq 0. \label{eq:dfn_positive_definite_maps_group}
\end{align}
for all subsets $h:=\{h_s\}_{s \in G}\subset\mathbb{C}^n$ whose elements are indexed by the elements of $G$. The positive definiteness of maps on a group is closely related to the positivity of their Fourier transforms via Bochner's theorem. Let $G$ be a finite group, and let $f:G \to \mathbb{C}$ be a complex valued function. Bochner's theorem states that $f$ is positive definite if and only if its Fourier transform $\widehat{f}(\rho)$ is positive semidefinite for all $\rho \in \mathrm{IrrU(G)}$, where $\mathrm{IrrU(G)}$ is the complete set of inequivalent irreducible unitary representations of the group $G$. A proof can be found in~\cite{DeCorte2014}. For matrix valued map $\Phi: G \to M_n(\mathbb{C})$, the above Bochner's theorem remains valid and can be stated explicitly as follows:
\\
\\
\textit{Bochner's theorem for matrix valued maps}: Let $G$ be a finite group. Then a map $\Phi: G \to M_n(\mathbb{C})$ is positive definite if and only if $\widehat{\Phi}(\rho)$ is positive semidefinite for all $\rho \in \rm{IrrU}(G)$, where $\rm{IrrU}(G)$ represents a complete set of inequivalent irreducible unitary representations of $G$.
\\
A proof is provided in Appendix~\ref{Bochner's theorem for matrix valued maps}. 
\\
\\
Now we are in a position to formulate and prove Bochner's theorem for finite inverse semigroups. We begin by defining positive definite maps on the contracted algebra of a finite inverse semigroup.
\begin{dfn}
    Let $S$ be a finite inverse semigroup with $0$ denoting the zero element, and let $\mathcal{H}=\mathbb{C}^n$ be a Hilbert space.  A linear map $\Phi: \mathbb{C}_0[S] \to \mathcal{B}(\mathcal{H})$ is called positive definite if the matrix $[\Phi(s^{-1}s')] \in \mathcal{B}(\mathcal{H}^{\bigoplus (|S|-1)})$ is positive semidefinite, i.e.,
\begin{align}
    \sum_{\substack{s,s' \in S \\ s\neq 0, s' \neq 0 }} \inner{\Phi(s^{-1}s')h_{s'},h_s} \geq 0 \label{eq:dfn_positive_definite_maps_inverse_semigroup}
\end{align}
for all subsets $h:=\{h_s\}_{s \in S \setminus\{0\} }\subset\mathbb{C}^n$ whose elements are indexed by the nonzero elements of $S$.
\end{dfn}
In the following, we present two alternative characterizations of positive definite maps (Proposition~\ref{prop:alternative characterization of positive definite maps 1} and Proposition~\ref{prop:alternative characterization of positive definite maps 2}), which will be used in the proof of Bochner's theorem for finite inverse semigroups.
\begin{proposition} \label{prop:alternative characterization of positive definite maps 1}  
    A linear map $\Phi: \mathbb{C}_0[S] \to \mathcal{B}(\mathcal{H})$ is positive definite if and only if the matrix $[\Phi(\floor{s^{-1}}\floor{s'})] \in \mathcal{B}(\mathcal{H}^{\bigoplus (|S|-1)})$ is positive semidefinite, \emph{i.e.,}
\begin{align}
    \sum_{\substack{s,s' \in S \\ s\neq 0, s' \neq 0 }} \inner{\Phi(\floor{s^{-1}}\floor{s'})h_{s'},h_s} \geq 0 \label{eq:dfn_positive_definite_maps_inverse_semigroup_1}
\end{align}
for all subsets $h:=\{h_s\}_{s \in S \setminus\{0\} }\subset\mathbb{C}^n$ whose elements are indexed by nonzero elements of $S$.
\end{proposition}
\begin{prof}
    We begin by recalling the definition of the groupoid basis $\floor{s}$ of the contracted algebra $\mathbb{C}_0[S]$ (see Eq.~(\ref{eq:groupoid_basis})).
    \begin{align}
        \lfloor s \rfloor = \sum_{\substack{t \in S \\ t \leq s}} \mu(t,s) t \in \mathbb{C}_{0}[S],
    \end{align}
    where $t \neq 0$. From this expression, it follows that
\begin{align}
        \lfloor s^{-1} \rfloor &= \sum_{\substack{t \in S \\ t \leq s}} \mu(t^{-1},s^{-1}) t^{-1}
        =\sum_{\substack{t \in S \\ t \leq s}} \mu(t^{-1}t,s^{-1}s) t^{-1}=\sum_{\substack{t \in S \\ t \leq s}} \mu(t,s) t^{-1}.
    \end{align}
    In the last two equalities we have used the fact that $\mu(t,s)=\mu(tt^{-1},ss^{-1})=\mu(t^{-1}t,s^{-1}s)$ (See \cite{STEINBERG2006866}). 
    Now, with $s,s' \neq 0$, we have
    \begin{align}
    \sum_{\substack{s,s' \in S }} \inner{\Phi(\floor{s^{-1}}\floor{s'})h_{s'},h_s}&= \sum_{\substack{s,s' \in S}}\sum_{\substack{t \in S \\ t \leq s}} \sum_{\substack{t' \in S \\ t' \leq s'}} \mu(t',s') \mu(t,s) \inner{\Phi(t^{-1}t')h_{s'},h_s} \nonumber\\
    &=\sum_{\substack{s,s',t,t' \in S \\ t \leq s\\t' \leq s'}} \mu(t',s') \mu(t,s) \inner{\Phi(t^{-1}t')h_{s'},h_s} \\
    &= \sum_{\substack{t,t' \in S}}   \inner{\Phi(t^{-1}t')\sum_{\substack{s' \in S \\ t' \leq s' }}\mu(t',s')h_{s'},\sum_{\substack{s\in S \\ t \leq s}}\mu(t,s)h_s}.
\end{align}
Now, by defining a family $h^0:=\{h^0_t\}$ consisting of elements $h^0_t:=\sum_{\substack{s\in S \\ t \leq s}}\mu(t,s)h_s$ (which can be inverted as $h_s=\sum_{\substack{t\in S \\ t \geq s}}h^0_t$), we can rewrite the above equation as
\begin{align}
    \sum_{\substack{s,s' \in S}} \inner{\Phi(\floor{s^{-1}}\floor{s'})h_{s'},h_s}
    &=\sum_{\substack{t,t' \in S}}   \inner{\Phi(t^{-1}t')h^0_{t'},h^0_t}.
\end{align}
This completes the proof.
\end{prof}
\\
\begin{proposition} \label{prop:alternative characterization of positive definite maps 2} 
     A linear map $\Phi: \mathbb{C}_0[S] \to \mathcal{B}(\mathcal{H})$ is positive definite if and only if the matrix $[\Phi(\floor{s^{-1}}\floor{s'})] \in \mathcal{B}(\mathcal{H}^{\bigoplus |D_k|})$ is positive semidefinite for all $\mathcal{D}$-classes $D_k$,  i.e.,
\begin{align}
    \sum_{\substack{s,s' \in D_k \\ s\neq 0, s' \neq 0 }} \inner{\Phi(\floor{s^{-1}}\floor{s'})h_{s'},h_s} \geq 0 \label{eq:dfn_positive_definite_maps_inverse_semigroup_2}
\end{align}
for all $\mathcal{D}$-classes $D_k$ and all subsets $h:=\{h_s\}_{s \in D_k }\subset\mathbb{C}^n$ whose elements are indexed by the elements of $D_k$.
\end{proposition}
\begin{prof}
    If $ss^{-1}=s's'^{-1}$, then $s$ and $s'$ belong to the same $\mathcal{D}$-class. Consequently, if $s$ and $s'$ lie in different $\mathcal{D}$-classes, then one has $ss^{-1}\neq s's'^{-1}$. The multiplication rule for the groupoid basis (See Eq.~(\ref{gropoid_basis_multiplication}) implies that $\floor{s^{-1}}\floor{s'}=0$ whenever $ss^{-1}\neq s's'^{-1}$. As a result, the matrix $[\Phi(\floor{s^{-1}}\floor{s'})]$ is block diagonal with the blocks indexed by the $\mathcal{D}$-classes of $S$. Therefore, $\Phi$ is positive definite if and only if, for every $\mathcal{D}$-classes $D_k$,
\begin{align}
    \sum_{\substack{s,s' \in D_k \\ s\neq 0, s' \neq 0 }} \inner{\Phi(\floor{s^{-1}}\floor{s'})h_{s'},h_s} \geq 0 \label{eq:dfn_positive_definite_maps_inverse_semigroup_2}
\end{align}
 for all subsets $h:=\{h_s\}_{s \in D_k }\subset\mathbb{C}^n$ whose elements are indexed by the elements of $D_k$. This completes the proof.
\end{prof}
\\
In the following proposition, we generalize the Schur orthogonality relation to finite inverse semigroups. This result plays a crucial role in the proof of Bochner's theorem for finite inverse semigroup.
\begin{proposition}[Schur Orthogonality relation for inverse semigroup] \label{Schur Orthogonality relation for inverse semigroup}
     Let $S$ be a finite inverse semigroup, and let $\{ D_k\}_{k=1}^n$ denote its $\mathcal{D}$-classes. Fix a $\mathcal{D}$-class $D_k$, and let $\overline{\rho}_k,\overline{\rho'}_k \in \mathrm{IrrU}(\mathbb{C}_0[S])$, where $\rm{IrrU}(\mathbb{C}_0[S])$ denotes a complete set of inequivalent irreducible representations of $\mathbb{C}_0[S]$ induced by a complete set of inequivalent irreducible unitary representations of the maximal subgroups of $S$, as described in Theorem~\ref{lemma:Stineberg_1}. Let $a,b,c,d \in D_k$ be idempotents. Then 
    \begin{align}
        \sum_{s \in D_k} ((\overline{\rho}_k(\floor{s}))_{a,b})_{i,j} \overline{((\overline{\rho'}_k(\floor{s}))_{c,d})_{k,l}}&=\begin{cases}
            \frac{r_k|G_{e_k}|}{d_{\overline{\rho}_k}} \delta_{a,c} \delta_{b,d} \delta_{i,k} \delta_{j,l} & \text{when}~ \overline{\rho}_k=\overline{\rho'}_k, \\
            0 & \text{when}~ \overline{\rho}_k \neq \overline{\rho'}_k,
        \end{cases} \label{eq:Schur orthogonal relations for D_k}
    \end{align}
    where $\overline{(\cdot)}$ denotes complex conjugation, and $((\overline{\rho}_k(\floor{s}))_{a,b})_{i,j}$ is the $(i,j)$-th entry of the $(a,b)$-th block of $\overline{\rho}_k(\floor{s})$.
\end{proposition}
\begin{prof}
    The action of the induced representation $\overline{\rho}_k$ on the groupoid basis element $\floor{s}$ is given by 
    \begin{align}
       \overline{\rho}_k(\floor{s})=E_{\rm{ran}(s),\rm{dom}(s)} \otimes \rho_{k}(p^{-1}_{\rm{ran}(s)}\, s \, p_{\rm{dom}(s)}) \in M_{r_k} (M_{n}(\mathbb{C})).\label{SS1}
    \end{align}
    Notice that $u:=p^{-1}_{\rm{ran}(s)}\, s \, p_{\rm{dom}(s)} \in G_{e_k}$. Now, from Eq.~(\ref{SS1}), it follows immediately that $$\overline{((\overline{\rho}_k(\floor{s}))_{a,b})_{i,j}}=\overline{(\rho_{k}(u))_{i,j}} \delta_{a, \mathrm{ran}(s)} \delta_{b, \mathrm{dom}(s)}.$$ Consequently,
    \begin{align}
       \scalemath{0.85}{ \sum_{s \in D_k} ((\overline{\rho}_k(\floor{s}))_{a,b})_{i,j} \overline{((\overline{\rho'}_k(\floor{s}))_{c,d})_{k,l}}= \sum_{s \in D_k} (\rho_{k}(u))_{i,j} \delta_{a, \mathrm{ran}(s)} \delta_{b, \mathrm{dom}(s)} \overline{(\rho'_{k}(u))_{k,l}} \delta_{c, \mathrm{ran}(s)} \delta_{d, \mathrm{dom}(s)}. }\label{SS2}
    \end{align}
    Next, observe that the map $\vartheta: D_k \to M_{r_k}(G_{e_k})$ defined by $$s \mapsto  (p^{-1}_{\rm{ran}(s)}\, s \, p_{\rm{dom}(s)})E_{\rm{ran}(s),\rm{dom}(s)}$$ is one-to-one correspondence with the inverse given by $$g E_{e,f} \mapsto p_{e} g p_{f}^{-1}.$$ Hence any $s \in D_k$ can be written as $s=p_{e} g p_{f}^{-1}$ for some pair of idempotents $(e,f) \in D_k$ and $g \in G_{e_k}$. It should be noted that $\mathrm{ran}(p_{e} g p_{f}^{-1})=e$ and $\mathrm{dom}(p_{e} g p_{f}^{-1})=f$ for all $g \in G_{e_k}$. A necessary condition for the terms in the summation of Eq.~(\ref{SS2}) to be non-vanishing is $a=\mathrm{ran}(s)$ and $b=\mathrm{dom}(s)$. So, the possible non-vanishing contributions come from summing over those $s \in D_k$ that have the form $s=p_{a} g p_{b}^{-1}$. With this observation, we have
    \begin{align}
        \sum_{s \in D_k} ((\overline{\rho}_k(\floor{s}))_{a,b})_{i,j} \overline{((\overline{\rho'}_k(\floor{s}))_{c,d})_{k,l}}&= \sum_{g \in G_{e_k}} (\rho_{k}(g))_{i,j}  \overline{(\rho'_{k}(g))_{k,l}} \delta_{c, a} \delta_{d, b}. \label{SS3}
    \end{align}
    Now, applying Schur orthogonality relation~\cite{serre1977linear,terras1999fourier,steinberg2012representation} for the inequivalent irreducible unitary representations $\rho_k$ and $\rho'_k$ of the maximal subgroup $G_{e_k}$ in Eq.~(\ref{SS3}), and using the fact that $d_{\bar{\rho}_k}=r_k d_{\rho_k}$, where $d_{\bar{\rho}_k}$ represents the dimension of the induced representation $\overline{\rho}_k$, we have 
     \begin{align}
        \sum_{s \in D_k} ((\overline{\rho}_k(\floor{s}))_{a,b})_{i,j} \overline{((\overline{\rho'}_k(\floor{s}))_{c,d})_{k,l}}&=\begin{cases}
            \frac{r_k|G_{e_k}|}{d_{\overline{\rho}_k}} \delta_{a,c} \delta_{b,d} \delta_{i,k} \delta_{j,l} & \text{when}~ \overline{\rho}_k=\overline{\rho'}_k, \\
            0 & \text{when}~ \overline{\rho}_k \neq \overline{\rho'}_k.
        \end{cases}
    \end{align}
    This completes the proof.
\end{prof}
\\

Before proving Bochner's theorem for finite inverse semigroups, we recall from Eq.~(\ref{eq:representation of C_{0}}) that for $s \in D_k$, the induced representation is given by $$\overline{\rho}_k(\floor{s})= E_{\rm{ran}(s),\rm{dom}(s)} \otimes \rho_{k}(p^{-1}_{\rm{ran}(s)}\, s \, p_{\rm{dom}(s)}),$$ while $\overline{\rho}_k(\floor{s})=0$ if $s \notin D_k$. Let $\rho_k$ be a unitary representation of the maximal subgroup $G_{e_k}$. Then by taking adjoint yields $$(\overline{\rho}_k(\floor{s}))^{\dagger}= E_{\rm{dom}(s),\rm{ran}(s)} \otimes \rho_{k}(p^{-1}_{\rm{dom}(s)}\, s^{-1} \, p_{\rm{ran}(s)}).$$ By Eq.~(\ref{eq:Induced representation action on inverse groupoid}), this can be written equivalently as $$(\overline{\rho}_k(\floor{s}))^{\dagger}=\overline{\rho}_k(\floor{s^{-1}}).$$ This will be used in proof Bochner's theorem for finite inverse semigroup.
\begin{theorem}[Bochner's theorem for finite inverse semigroups]\label{thm:Bochner's theorem for finite inverse semigroups}
     Let $S$ be a finite inverse semigroup, and let $\Phi: \mathbb{C}_0[S] \to M_n(\mathbb{C})$ be a linear map. Then the linear map $\widetilde{\Phi}:\mathbb{C}_0[S] \to M_n(\mathbb{C})$ defined by $\widetilde{\Phi} (\floor{s})=\sum_{\substack{t \in S \\ t \geq s}} \Phi (t)$ is positive definite if and only if the Fourier transform $\widehat{\Phi}(\overline{\rho}_k)$ is positive semidefinite for all $\overline{\rho}_k \in \rm{IrrU}(\mathbb{C}_0[S])$, where $\rm{IrrU}(\mathbb{C}_0[S])$ denotes the complete set of inequivalent irreducible representations of $\mathbb{C}_0[S]$ induced by a complete set of inequivalent irreducible unitary representations of the maximal subgroups of $S$.
\end{theorem}
\begin{prof}
We begin by recalling the inversion formula for finite inverse semigroups. For $s \in D_k$, it is given by
\begin{align}
    \widetilde{\Phi}(\floor{s})&=\frac{1}{r_k|G_{e_k}|} \sum_{\sigma \in \rm{IrrU}(\mathbb{C}_{0}[S])} d_{\sigma} \tr_1  \big[(\sigma(\floor{s^{-1}}) \otimes \mathbb{I}) \widehat{\Phi}(\sigma)\big].
\end{align}
    Since $\sigma \in \rm{IrrU}(\mathbb{C}_{0}[S])$, we have $\sigma(\floor{s^{-1}})=(\sigma(\floor{s}))^{\dagger}$. Therefore, the inversion formula can be rewritten as
     \begin{align}
    \widetilde{\Phi}(\floor{s})&=\frac{1}{r_k|G_{e_k}|} \sum_{\sigma \in \rm{IrrU}(\mathbb{C}_{0}[S])} d_{\sigma} \tr_1  \big[((\sigma(\floor{s}))^{\dagger} \otimes \mathbb{I}) \widehat{\Phi}(\sigma)\big]. \label{eq:FI1}
\end{align}
  Let $s,t \in D_k$ be such that $s^{-1}s=tt^{-1}$. For such a pair of $s$ and $t$, it follows from Eq.~(\ref{eq:FI1}) that
   \begin{align}
    \widetilde{\Phi}(\floor{s}\floor{t})&=\frac{1}{r_k|G_{e_k}|} \sum_{\sigma \in \rm{IrrU}(\mathbb{C}_{0}[S])} d_{\sigma} \tr_1  \big[(\sigma(\floor{t})^{\dagger} \sigma(\floor{s})^{\dagger} \otimes \mathbb{I}) \widehat{\Phi}(\sigma)\big].
\end{align}
Notice that the above equation remains valid even when $s^{-1}s\neq tt^{-1}$. Because in this case $\floor{s}\floor{t}=0$, and $\widetilde{\Phi}$ and $\sigma$ are being linear, we have $\widetilde{\Phi}(\floor{s}\floor{t})=\sigma(\floor{s}\floor{t})=0$. Now, for  vectors $h_s \in \mathbb{C}^n$, indexed by $s \in D_k$, we have 
 \begin{align}
    \sum_{s,t \in D_k}\inner{\widetilde{\Phi}(\floor{s^{-1}}\floor{t})h_t, h_s}&=\sum_{s,t \in D_k}\tr \left( \widetilde{\Phi}(\floor{s^{-1}}\floor{t}) h_t h_s^{\dagger}\right) \nonumber\\
    &=\frac{1}{r_k|G_{e_k}|}\!\! \sum_{s,t \in D_k}\sum_{\sigma \in \rm{IrrU}(\mathbb{C}_{0}[S])} \!\!\!\!\!\!\!\!\!\!\!d_{\sigma} \tr  \big[(\sigma(\floor{t})^{\dagger} \sigma(\floor{s})\otimes h_t h_s^{\dagger}) \widehat{\Phi}(\sigma)\big] \nonumber\\
    &=\frac{1}{r_k|G_{e_k}|}\sum_{\sigma \in \rm{IrrU}(\mathbb{C}_{0}[S])} d_{\sigma} \tr  \big[A_k(\sigma,h)^{\dagger}A_k(\sigma,h) \widehat{\Phi}(\sigma)\big], \label{eq:Bochner inverse semigroup}
\end{align}
where 
\begin{align}
    A_k(\sigma,h):=\sum_{s \in D_k} \sigma(\floor{s}) \otimes h_s^{\dagger},
\end{align}
 with $h:=\{h_s \}_{s \in D_k}$. Observe that Eq.~(\ref{eq:Bochner inverse semigroup}) is true for all $\mathcal{D}$-classes $D_k$. Now, assume that $\widehat{\Phi}(\sigma)$ is positive semidefinite for all $\sigma \in \rm{IrrU}(\mathbb{C}_{0}[S])$. Since $A_k(\sigma,h)^{\dagger}A_k(\sigma,h)$ is positive semidefinite for every choice of $D_k$, $\sigma$ and $h$, each term in the right hand side of Eq.~(\ref{eq:Bochner inverse semigroup}) is non-negative. Consequently, $$\sum_{s,t \in D_k}\inner{\widetilde{\Phi}(\floor{s^{-1}}\floor{t})h_t, h_s} \geq 0$$ for all $\mathcal{D}$-classes $D_k$. By Proposition~\ref{prop:alternative characterization of positive definite maps 2}, this implies that the map $\widetilde{\Phi}$ is positive definite. This completes the proof in one direction.

For the other direction, fix $\overline{\rho}_k^0 \in \rm{IrrU}(\mathbb{C}_{0}[S])$ induced by $\rho^0_{k} \in \rm{IrrU}(G_{e_{k}})$ and choose $h^{0}$ as the subset consisting of the vectors $h^0_s$ of the form
\begin{align}
h^{0}_s= \tr_{d_{\overline{\rho}_k^0 }} \left(\overline{\rho}^0_k(\floor{s}) \otimes \mathbb{I}) B^{\dagger}\right),
\end{align} 
where $B \in M_{d_{\overline{\rho}^0_{k}}}(\mathbb{C}) \otimes \mathbb{C}^{n*}$, $\tr_{d_{\overline{\rho}^0_{k}}}$ denotes the trace over the space $M_{d_{\overline{\rho}^0_{k}}}(\mathbb{C})$ and $\mathbb{C}^{n*}$ is the dual space of $\mathbb{C}^n$. Writing $B$ as a finite sum $B=\sum_{p}X_p \otimes x_p$, where $X_p \in M_{d_{\overline{\rho}^0_{k}}}(\mathbb{C})$ and $x_p \in \mathbb{C}^{n*}$, we have
\begin{align}
    (h_s^0)^{\dagger}&=\tr_{d_{\overline{\rho}^0_{k}}} \left((\overline{\rho}^0_{k}\left( \floor{s} \right)^{\dagger} \otimes \mathbb{I})B\right)\\
    &=\sum_{c,d,p,k,l}\overline{((\overline{\rho}^0_{k}\left( \floor{s} \right))_{dc})_{lk}} (((X_p))_{dc})_{lk} x_p.
\end{align}
Here, $\overline{((\overline{\rho}^0_{k}\left( \floor{s} \right))_{dc})_{lk}}$ denotes the complex conjugate of the $(l,k)$-th matrix element of the $(d,c)$-th block of the block matrix $\overline{\rho}^0_k(\floor{s})$.
Now taking $\sigma=\overline{\rho}_k$, we have
\begin{align}
    A_k(\overline{\rho}_k,h^{0}) &=\sum_{s \in D_k} \overline{\rho}_k(\floor{s}) \otimes (h^0_s)^{\dagger} \nonumber\\
    &= \sum_{a,b,i,j}\sum_{s \in D_k} ((\overline{\rho}_{k}(\floor{s}))_{a,b})_{i,j} E_{ab} \otimes e_{ij} \otimes \sum_{c,d,p,k,l}\overline{((\overline{\rho}^0_{k}\left( \floor{s} \right))_{dc})_{lk}} (((X_p))_{dc})_{lk} x_p \nonumber\\
     &= \!\!\!\sum_{a,b,i,j}\sum_{c,d,p,k,l}\!\Bigg(\sum_{s \in D_k}\!\overline{((\overline{\rho}^0_{k}\left( \floor{s} \right))_{dc})_{lk}}((\overline{\rho}_{k}(\floor{s}))_{a,b})_{i,j}\Bigg) ((X_p)_{dc})_{lk}  E_{ab} \otimes e_{ij} \otimes  x_p \nonumber\\
      &=\begin{cases}
       \frac{r_k|G_{e_{k}}|}{d_{\overline{\rho}_k^0}} B & \text{when}~\overline{\rho}_k=\overline{\rho}_k^0, \\
       0 & \text{when}~\overline{\rho}_k \neq \overline{\rho}_k^0.
    \end{cases}
\end{align}
The last equality follows from the Schur orthogonality relation for finite inverse semigroups, Eq.~(\ref{eq:Schur orthogonal relations for D_k}).
Substituting this expression of $A_k(\overline{\rho}_k,h^{0})$ into Eq.~(\ref{eq:Bochner inverse semigroup}), we obtain
\begin{align}
    \sum_{s,t \in D_k}\inner{\widetilde{\Phi}(\floor{s^{-1}}\floor{t})h^0_t, h^0_s}
    &=\frac{1}{r_k|G_{e_k}|}\sum_{\sigma \in \rm{IrrU}(\mathbb{C}_{0}[S])} d_{\sigma} \tr  \big[A_k(\sigma,h^0)^{\dagger}A_k(\sigma,h^0) \widehat{\Phi}(\sigma)\big] \nonumber\\
    &=\frac{1}{r_k|G_{e_k}|} d_{\overline{\rho}_k^0} \tr  \big[A_k(\overline{\rho}_k^0,h^0)^{\dagger} A_k(\overline{\rho}_k^0,h^0)\widehat{\Phi}(\overline{\rho}_k^0)\big] \nonumber\\
    &=\frac{d_{\overline{\rho}_k^0}}{r_k|G_{e_k}|} \left(\frac{r_k|G_{e_{k}}|}{d_{\overline{\rho}_k^0}}\right)^2 \tr  \big[B^{\dagger}B\widehat{\Phi}(\overline{\rho}_k^0)\big]\nonumber\\
    &= \frac{r_k|G_{e_{k}}|}{d_{\overline{\rho}_k^0}} \tr  \big[B^{\dagger}B\widehat{\Phi}(\overline{\rho}_k^0)\big].
\end{align}
Now assume that $\widetilde{\Phi}:\mathbb{C}_0[S] \to M_n(\mathbb{C})$ is positive definite, then the left hand side of the above equation is non-negative. Since $B$ and $\overline{\rho}_k^0$ were chosen arbitrarily, it follows that $\widehat{\Phi}(\overline{\rho}_k)$ is positive semidefinite for all $\overline{\rho}_k \in \rm{IrrU}(\mathbb{C}_0[S])$. This completes the proof.
\end{prof}
\\

The Stinespring dilation theorem for completely positive maps is a central result in $C^{*}$-algebra. The Stinespring dilation theorem  also exists for positive definite maps from a group to a von Neumann algebra~\cite{kasparov1980hilbert,Ozawa}. Here we prove an analogous dilation theorem for positive definite maps on the contracted algebra $\mathbb{C}_0[S]$ of a finite inverse semigroup $S$.
\begin{theorem}[Stinespring Dilation]\label{thm:Stinespring Dilation theorem for inverse semigroup}
    Let $S$ be a finite inverse semigroup and $\mathbb{C}_0[S]$ the the contracted algebra of $S$. Let $\Phi: \mathbb{C}_0[S] \to M_n(\mathbb{C})$ be a linear map. Then $\Phi$ is positive definite if and only if there exists a Hilbert Space $\mathcal{H}$, a bounded operator $V:\mathbb{C}^n \to \mathcal{H}$ and a $*$-homomorphism $\pi:\mathbb{C}_0[S] \to \mathcal{B}(\mathcal{H})$ such that for all $s \neq 0$
    \begin{align}
        \Phi(\floor{s})=V^{*}\pi(\floor{s})V,
    \end{align}
    with
    \begin{align}
        V^{*}V=\Phi(\mathds{1}),
    \end{align}
    where $\mathds{1}:=\sum_{e \in E(S)} \floor{e}$ is the identity of the contracted algebra $\mathbb{C}_0[S]$, and $E(S)$ denotes the set of non-zero idempotents of $S$.
\end{theorem}
\begin{prof}
The proof is essentially based on GNS construction. We consider the space $\mathbb{C}_0[S] \otimes \mathbb{C}^n$. Any generic element in $\mathbb{C}_0[S] \otimes \mathbb{C}^n$ can be written as $f=\sum_{\substack{s \in S\\s \neq 0}} \floor{s} \otimes f_s$, where $f_s \in \mathbb{C}^n$. Assuming that the map $\Phi$ is positive definite, we can define a positive semidefinite sesquilinear form on $\mathbb{C}_0[S] \otimes \mathbb{C}^n$ as follows:
\begin{align}
    \inner{f,h}:=\sum_{\substack{s,t \in S\\s,t \neq 0}} \inner{\Phi(\floor{s^{-1}}\floor{t})f_t,h_s}.
\end{align}
Let us define the set $\mathcal{N}:=\{f \in \mathbb{C}_0[S] \otimes \mathbb{C}^n: \inner{f,f}=0 \}$. Note that $\mathcal{N}$ is a subspace. Now we consider the quotient space $\mathbb{C}_0[S] \otimes \mathbb{C}^n/\mathcal{N}$ whose elements are the equivalence classes $[f]$ where the equivalence relation is defined as:$f \equiv f'$ if and only if $f-f' \in \mathcal{N}$. Now one can define an inner product on the quotient space $\mathbb{C}_0[S] \otimes \mathbb{C}^n/\mathcal{N}$ as follows:
\begin{align}
    \inner{[f],[h]}:=\sum_{\substack{s,t \in S\\s,t \neq 0}} \inner{\Phi(\floor{s^{-1}}\floor{t})f_t,h_s}.
\end{align}
It can be shown that the above defined inner product is well defined. With this inner product defined, the quotient space becomes an inner product space, and since we are working on finite dimensions, it is a Hilbert space which we denote by $\mathcal{H}:=\mathbb{C}_0[S]\otimes \mathbb{C}^n/\mathcal{N}$.
Now, define a linear map on $\mathcal{H}$ as
\begin{align}
    \pi(\floor{s})[f]:=[\floor{s}\cdot f],
\end{align}
where the element $\floor{s} \cdot f \in \mathbb{C}_0[S] \otimes \mathbb{C}^n$ is defined by 
\begin{align}
    \floor{s} \cdot f:=\sum_{\substack{t \in S\\t \neq 0}} \floor{s}\floor{t} \otimes f_t.
\end{align}
The map $\pi$ extends linearly to a map $\pi: \mathbb{C}_0[S] \to \mathcal{B}(\mathcal{H})$. One can simply verify that 
\begin{align}
   \pi(\floor{s}\floor{t})=\pi(\floor{s})\pi(\floor{t}). 
\end{align}
Note that, due to the linearity of $\pi$, the relation $\pi(\floor{s}\floor{t})=\pi(\floor{s})\pi(\floor{t})$ is consistent with the fact that $\floor{s}\floor{t}=0$ whenever $\mathrm{dom(s)} \neq \mathrm{ran}(t)$. Define another map $V:\mathbb{C}^n \to \mathcal{H}$ by
\begin{align}
    V(x):=\left[ \mathds{1} \otimes x \right],
\end{align}
where $\mathds{1}:=\sum_{e \in E(S)} \floor{e}$ is the identity of the contracted algebra $\mathbb{C}_0[S]$ (see~\cite{STEINBERG20081521}). Let $V^{*}:\mathcal{H} \to \mathbb{C}^n$ denote the adjoint of $V$. Then we have 
\begin{align}
    \inner{V^{*}[f],x}&=\inner{[f],\left[ \mathds{1} \otimes x \right]} \nonumber\\
    &=\inner{[f],\left[ \sum_{e \in E(S)} \floor{e} \otimes x \right]} \nonumber\\
    &=\sum_{\substack{t \in S\\t \neq 0}} \sum_{e \in E(S)}\inner{\Phi(\floor{e^{-1}}\floor{t})f_t,x}\nonumber\\
    &=\sum_{\substack{t \in S\\t \neq 0}} \inner{\Phi\left(\sum_{e \in E(S)}\floor{e}\floor{t}\right)f_t,x}\nonumber\\
    &=\sum_{\substack{t \in S\\t \neq 0}} \inner{\Phi\left(\floor{t}\right)f_t,x}.
\end{align}
%where in the last equality we have used the fact that $\mathbb{C}_0[S]$ is a unital algebra with the identity element being $\mathds{1}:=\sum_{e \in E(S)}\floor{e}$ . 
Hence, the adjoint map $V^{*}$ is given by 
\begin{align}
    V^{*}[f]&=\sum_{\substack{t \in S\\t \neq 0}} \Phi\left(\floor{t}\right)f_t.
\end{align}
Now, using the definition of $V$ and the action of $V^*$ as provided above, we obtain
\begin{align}
    V^{*}\pi(\floor{s})V(x)=V^{*}\pi(\floor{s})\left[\mathds{1} \otimes x\right]=V^{*}\left[ \floor{s} \otimes x \right]=\Phi(\floor{s})x.
\end{align}
Thus 
\begin{align}
    \Phi(\floor{s})=V^{*}\pi(\floor{s})V
\end{align}
Moreover,
\begin{align}
    V^{*}V(x)&=V^{*}\left[\mathds{1} \otimes x\right] 
    =\Phi\left(\sum_{e \in E(S)} \floor{e}\right)x 
    =\Phi\left(\mathds{1}\right)x.
\end{align}
So,
\begin{align}
    V^{*}V=\Phi\left(\mathds{1}\right).
\end{align}
Now, we show that $\pi$ is a $*$-homomorphism. For $u \in S$ and $[f], [h] \in \mathcal{H}$, we compute the following.
\begin{align}
    \inner{\pi(\floor{u})^{\dagger}[f],[h]}&= \inner{[f],\pi(\floor{u})[h]}\nonumber\\
    &= \inner{[f],[\floor{u}\cdot h]} \nonumber\\
    &= \inner{\left[\sum_{\substack{t \in S\\t \neq 0}} \floor{t} \otimes f_t\right],\left[\sum_{\substack{s \in S\\s \neq 0}} \floor{u}\floor{s} \otimes h_s\right]}\nonumber\\
    &=\sum_{\substack{s,t \in S\\s,t \neq 0}} \inner{\Phi(\floor{s^{-1}}\floor{u^{-1}}\floor{t})f_t,h_s}\nonumber\\
    &= \inner{\left[\sum_{\substack{t \in S\\t \neq 0}} \floor{u^{-1}}\floor{t} \otimes f_t\right],\left[\sum_{\substack{s \in S\\s \neq 0}}\floor{s} \otimes h_s\right]}\nonumber\\
    &=\inner{\pi(\floor{u^{-1}})[f],[h]}.
\end{align}
Hence,
\begin{align}
    \pi(\floor{u})^{\dagger}=\pi(\floor{u^{-1}})
\end{align}
 for all $u \in S$, which implies that $\pi$ is a $*$-homomorphism.
\\

Conversely, suppose that a linear map $\Phi: \mathbb{C}_0[S] \to M_n(\mathbb{C})$ can be written as $ \Phi(\floor{s})=V^{*}\pi(\floor{s})V$ for some  Hilbert Space $\mathcal{H}$, a bounded operator $V:\mathbb{C}^n \to \mathcal{H}$ and a $*$-homomorphism $\pi:\mathbb{C}_0[S] \to \mathcal{B}(\mathcal{H})$. Then for any subset $\{ h_s\}_{s \in S \setminus\{ 0\}} \subset \mathbb{C}^{n}$, we have 
    \begin{align}
         \sum_{\substack{s,t \in S \\ s\neq 0, t \neq 0 }} \inner{\Phi(\floor{s^{-1}}\floor{t})h_t,h_s} &=\sum_{\substack{s,t \in S \\ s\neq 0, t \neq 0 }} \inner{V^{*}\pi(\floor{s^{-1}})\pi(\floor{t})Vh_t,h_s} \nonumber\\
         &=\sum_{\substack{s,t \in S \\ s\neq 0, t \neq 0 }} \inner{\pi(\floor{t})Vh_t,\pi(\floor{s^{-1}})^{\dagger}Vh_s}\nonumber\\
         &= \inner{\sum_{\substack{t \in S \\ t \neq 0 }}\pi(\floor{t})Vh_t,\sum_{\substack{s \in S \\ s\neq 0 }}\pi(\floor{s})Vh_s} \label{SS10}\\
         & \geq 0, \nonumber
    \end{align}
    where in Eq~(\ref{SS10}) we have used the fact that $\pi$ is a $*$-homomorphism. This shows that $\Phi$ is a positive definite map. This completes the proof.
\end{prof}
\subsection{Choi’s Theorem as a Special Case of Bochner’s Theorem} \label{subsec:Choi theorem on Completely positive maps and Bochner's theorem on positive definite maps on finite dimensional matrix algebra}
As discussed earlier, the matrix algebra $M_n(\mathbb{C})$ is a contracted algebra of the inverse semigroup $S_M=\{e_{ij}\}_{i,j=1}^n \cup \{0\}$. Therefore, it follows from Theorem~\ref{thm:Stinespring Dilation theorem for inverse semigroup} that a linear map $\Phi:M_m(\mathbb{C}) \to M_n(\mathbb{C})$ is positive definite if and only if it completely positive.
On the other hand, Bochner's Theorem~(\ref{thm:Bochner's theorem for finite inverse semigroups}) for finite inverse semigroup implies that a linear map  $\Phi:M_m(\mathbb{C}) \to M_n(\mathbb{C})$ is positive definite if and only if its Fourier transform $\widehat{\Phi}(\overline{\rho}_k)$ is positive semidefinite for all $\overline{\rho}_k \in \mathrm{IrrU}(M_m(\mathbb{C}))$.

From Proposition~\ref{prop:1}, the inverse semigroup $S_M$ has exactly two $\mathcal{D}$-classes, namely $$D_0=\{0\} \quad\text{and} \quad D_1=\{e_{ij}\}_{i,j=1}^n,$$ with $D_1$ being the only nontrivial class. Let us fix the idempotent $e_{11}$. Now, the maximal subgroup at $e_{11}$ is $G_{e_{11}}:=\{e_{11}\}$. Moreover, a straightforward computation shows $$\floor{e_{ij}}=e_{ij},\quad  \mathrm{ran}(e_{ij})=e_{ii}, \quad  \mathrm{dom}(e_{ij})=e_{jj},$$ $$p_{\mathrm{ran}(e_{ij})}=e_{i1}, \quad and \quad  p_{\mathrm{dom}(e_{ij})}=e_{j1}.$$ Now from the definition of the induced representation $\overline{\rho}_k$~(see Theorem~\ref{lemma:Stineberg_1}), we obtain
\begin{align}
    \overline{\rho}_1(e_{ij})&=E_{e_{ii},e_{jj}} \otimes \rho_{1}(e_{1i}e_{ij}e_{j1})\nonumber\\
    &=E_{e_{ii},e_{jj}} \otimes \rho_{1}(e_{11}). \nonumber
\end{align}
Since $\rho_{1}$ is an irreducible unitary representation of the singleton group $G_{e_{11}}=\{e_{11}\}$, we have $\rho_{1}(e_{11})=1$. Identifying $E_{e_{ii},e_{jj}}$ with $e_{ij}$, we have $\overline{\rho}_1(e_{ij})=e_{ij}$. Now, the definition of Fourier transform of $\Phi$ (see Eq.~(\ref{eq:Fourier transform wrt induced representation_0})) yields the following,
\begin{align}
    \widehat{\Phi}(\overline{\rho}_1)&=\sum_{s \in S_M, s\neq 0} \overline{\rho}_1(\floor{s}) \otimes \widetilde{\Phi}(\floor{s})\nonumber\\
    &=\sum_{i,j=1}^n \overline{\rho}_1(e_{ij}) \otimes \Phi(e_{ij}) \label{SS11}\\
    &=\sum_{i,j=1}^n e_{ij} \otimes \Phi(e_{ij}), \label{SS12}
\end{align}
where in Eq.~(\ref{SS11}) we have used the fact that $\widetilde{\Phi}(\floor{e_{ij}})=\Phi(e_{ij})$. From Eq.~(\ref{SS12}) we see that $\widehat{\Phi}(\overline{\rho}_1)$ is precisely the Choi matrix of the linear map $\Phi$. Thus, in the context of finite dimensional matrix algebra, Bochner's theorem reduces to the Choi theorem on completely positive maps.
\subsection{Representations Characterizing Complete Positivity via Fourier Positivity
} \label{subsec:Relation between complete positivity of maps and positivity of their Fourier transforms}
In Section~\ref{subsec:Choi theorem on Completely positive maps and Bochner's theorem on positive definite maps on finite dimensional matrix algebra}, we showed that, for matrix algebras, Bochner's theorem on positive definite maps reduces to the Choi theorem for completely positive (CP) maps. It should be noted that the Fourier transforms that appeared in Bochner's theorem (Theorem~\ref{thm:Bochner's theorem for finite inverse semigroups}) are taken with respect to the induced representations arising from the irreducible unitary representations of the maximal subgroups of the finite inverse semigroup. This naturally raises the question of what happens to the CP-positivity correspondence when a general representation of the matrix algebra is considered. More precisely, one may ask the following.
\\
\\
Let $\Phi: M_m(\mathbb{C}) \to M_n(\mathbb{C})$ be a completely positive map and let $\rho:M_m(\mathbb{C}) \to M_{d_{\rho}}(\mathbb{C})$ be a representation. Then, is $\Phi$ completely positive if and only if its Fourier transform $\widehat{\Phi}(\rho)$ is positive semidefinite?
\\
\\
In the following theorem we provide a necessary and sufficient condition on the representation $\rho$ under which the answer to this question is affirmative.
\begin{theorem} \label{thm:CP vs positivity}
     Let $\Phi: M_m(\mathbb{C}) \to M_n(\mathbb{C})$ be a linear map and let $\rho:M_m \to M_{d_{\rho}}(\mathbb{C})$ be a representation. The correspondence between the complete positivity of $\Phi$ and the positivity of its Fourier transform $\widehat{\Phi}(\rho)$ holds if and only if $\rho$ is of the form $\rho(X)=U X U^{\dagger}$ for all $X \in M_m(\mathbb{C})$, where $U:\mathbb{C}^m \to \mathbb{C}^{d_{\rho}}$ is unitary with $d_{\rho}=m$.
\end{theorem}
\begin{prof}
    From the definition, we have 
    \begin{align}
        \widehat{\Phi}(\rho)&=\sum_{i,j} \rho(e_{ij}) \otimes \Phi(e_{ij}) \nonumber\\
        &=(\mathrm{id} \otimes \Phi)\sum_{i,j} \rho(e_{ij}) \otimes e_{ij}. 
    \end{align}
Now it can be shown that $$\sum_{i,j} \rho(e_{ij}) \otimes e_{ij}=\tau(C_{\rho^{*}}),$$ where $\rho^{*}$ is the adjoint of the representation $\rho:M_m(\mathbb{C}) \to M_{d_{\rho}}(\mathbb{C})$, $C_{\rho^{*}}$ is the Choi matrix of $\rho^{*}$, and $\tau$ denotes matrix transposition. Consequently,
\begin{align}
        \widehat{\Phi}(\rho)
        &=(\mathrm{id} \otimes \Phi)\tau(C_{\rho^{*}})=(\tau \otimes \Phi \circ \tau)C_{\rho^{*}} 
    \end{align}
which is equivalent to 
\begin{align}
    \tau(\widehat{\Phi}(\rho)) = (\mathrm{id} \otimes \tau \circ \Phi \circ \tau)C_{\rho^{*}}.
\end{align}
Observe that $\tau(\widehat{\Phi}(\rho))$ is positive semidefinite if and only if $\widehat{\Phi}(\rho)$ is positive semidefinite, and $\Phi$ is CP if and only if $\tau \circ \Phi \circ \tau$ is CP. Now, by Theorem~\ref{thm:Kye-Paulsen}, the CP-positivity correspondence holds if and only if $\rho^{*}$ is a complete order isomorphism, equivalently if and only if $\rho$ is a complete order isomorphism. Since $\rho$ is a representation, it must be of the form $\rho(X)=U X U^{\dagger}$ for all $X \in M_m(\mathbb{C})$, where $U:\mathbb{C}^m \to \mathbb{C}^{d_{\rho}}$ is unitary with $d_{\rho}=m$. This completes the proof.
\end{prof}
\section{Conclusion} \label{sec:Conclusion}
Bochner's theorem, one of the most celebrated theorems in harmonic analysis, provides a precise correspondence between positive definite maps and the positivity of their Fourier transforms. The central contribution of this work is the extension of this correspondence to the setting of finite inverse semigroups.

We establish a Bochner-type theorem for matrix-valued maps on the contracted algebra of a finite inverse semigroup. Exploiting the intrinsic partial order of inverse semigroups, positivity naturally arises at the level of Möbius transformed map. Our main result characterizes the positive definiteness of this transformed map in terms of the positivity of the Fourier transform of the original map with respect to a complete family of inequivalent irreducible representations of the contracted algebra induced by a complete family of inequivalent irreducible unitary representations of the maximal subgroups of the inverse semigroup. 

The proof of this Bochner-type theorem is supported by several structural results developed here for finite inverse semigroups. These include a convolution algebra structure on the space of linear maps, a Fourier transform and inversion formula, a Plancherel formula, alternative characterizations of positive definite maps, and a Schur orthogonality relation adapted to the inverse semigroup setting.

Within this framework, Choi's theorem on completely positive maps arises naturally as a special case when the inverse semigroup of matrix units are considered. Moreover, the Fourier transform with respect to identity representation coincides with the Choi matrix of the linear map, and the Fourier inversion formula reduces to the Choi inversion formula. This shows that Choi's characterization of complete positivity is a manifestation of a more general Bochner-type characterization of positive definite maps. 

%\section*{Declarations}

%\noindent\textbf{Conflict of Interest.}~~The authors have no conflict of interest to disclose.\\

%\bmhead{Data Availability Statement} No datasets were analyzed or generated, as this study is purely theoretical and mathematical.
\appendix
\section{Proof of Plancherel Formula for Finite Groups} \label{appendix:Proof of Plancherel formula for finite groups}
Let $G$ be a finite group, and let $\Phi, \Psi: G \to M_n(\mathbb{C})$ be linear maps. Using the Fourier inversion formula for maps on $G$, we have 
\begin{align}
     \Phi(g^{-1}) &= \frac{1}{|G|}\sum_{\rho \in \mathrm{Irr}(G)} d_{\rho} \tr_{d_{\rho}} \big[(\rho (g) \otimes \mathbb{I}) \widehat{\Phi}(\rho) \big].
\end{align}
Substituting this expression into $\sum_{g \in G} \Phi(g^{-1}) \Psi(g)$ and rearranging the terms yields
\begin{align}
\nonumber
    \sum_{g \in G} \Phi(g^{-1}) \Psi(g) &=   
     \frac{1}{|G|}\sum_{g \in G} \sum_{\rho \in \mathrm{Irr}(G)} d_{\rho} \tr_{d_{\rho}} \big[(\rho (g) \otimes \mathbb{I}) \widehat{\Phi}(\rho) (\mathbb{I} \otimes \Psi(g)) \big] \\
\nonumber    
    &=  \frac{1}{|G|}\sum_{g \in G} \sum_{\rho \in \mathrm{Irr}(G)} d_{\rho} \tr_{d_{\rho}} \big[ \widehat{\Phi}(\rho) (\mathbb{I} \otimes \Psi(g))(\rho (g) \otimes \mathbb{I}) \big] \\
\nonumber    
    &=  \frac{1}{|G|}\sum_{g \in G} \sum_{\rho \in \mathrm{Irr}(G)} d_{\rho} \tr_{d_{\rho}} \big[ \widehat{\Phi}(\rho) (\rho (g) \otimes \Psi(g)) \big] \\
\nonumber    
    &= \frac{1}{|G|}  \sum_{\rho \in \mathrm{Irr}(G)} d_{\rho} \tr_{d_{\rho}} \big[ \widehat{\Phi}(\rho) \widehat{\Psi}(\rho)  \big].
\end{align}
This completes the proof.
\section{Proof of Theorem~\ref{group_convolution}} \label{Appendix:Proof of Theorem A}
     A  map $\mathrm{\Phi} : G \rightarrow \widetilde{\mathcal{A}}$  can be completely specified by the tuple $(\mathrm{\Phi}(g_1), \mathrm{\Phi}(g_2),....,\mathrm{\Phi}(g_d) )$. Let  $L(G,\widetilde{\mathcal{A}})$ denote the set of all such maps. This is a vector space over $\mathbb{C}$ under the usual addition and scalar multiplication. Now a map $\mathrm{\Phi} \in L(G,\widetilde{\mathcal{A}})$ can be extended to the map $\mathrm{\Phi}: \mathbb{C}[G] \rightarrow \widetilde{\mathcal{A}}$ by linearity. The spaces $L(G,\widetilde{\mathcal{A}})=L(\mathbb{C}[G],\widetilde{\mathcal{A}})$ and $\mathbb{C}[G] \otimes \widetilde{\mathcal{A}}$ are isomorphic as vector spaces under the identification $(\mathrm{\Phi}(g_1), \mathrm{\Phi}(g_2),....,\mathrm{\Phi}(g_d) ) \leftrightarrow \sum_i g_i \otimes \mathrm{\Phi}(g_i)$ or equivalently by $\mathrm{\Phi} \leftrightarrow \sum_i g_i \otimes \mathrm{\Phi}(g_i)$. The algebra $\mathbb{C}[G] \otimes \widetilde{\mathcal{A}}$ has the multiplication rule induced by group multiplication in $G$ and multiplication rule in $\widetilde{\mathcal{A}}$. Let $a$ and $a'$ be two arbitrary elements in $\mathbb{C}[G] \otimes \widetilde{\mathcal{A}}$ and let $\mathrm{\Phi}$ and $\mathrm{\Phi}'$ be the corresponding maps in $L(\mathbb{C}[G],\widetilde{\mathcal{A}})$. Now we have
     \begin{align}
     \nonumber
         a \cdot a' &= \bigg(\sum_i g_i \otimes \mathrm{\Phi}(g_i)\bigg) \cdot \bigg( \sum_j g_j \otimes \mathrm{\Phi}'(g_j)\bigg) \\
         &= \sum_{i,j} g_{i}g_{j} \otimes \mathrm{\Phi}(g_i) \mathrm{\Phi}'(g_j). \label{group_table}
     \end{align}
     Now, in the group multiplication table, along a row or column, each element of the group appears once. So, it is guaranteed that the terms $g_{i}g_{j}$ for all possible values of $i$ and $j$ produce all elements of the group. We collect together all the terms in the summation of Eq.~(\ref{group_table}) for which $g_{i}g_{j}$ is equal to a particular group element, say $g_{k}$ and rewrite the summation as follows:
     \begin{align}
         a \cdot a' &= \sum_{k} g_{k} \otimes \bigg( \sum_{\substack{p,q: \\ g_p g_q=g_k}}\mathrm{\Phi}(g_p)\mathrm{\Phi}'(g_q) \bigg), \label{group_table_1}
     \end{align}
    where the summation inside the big bracket is over those pair of elements $(g_{p}, g_{q})$ such that $g_{p} g_{q}= g_{k}$. Note that the term inside the bracket is nothing but the convolution of the maps $\mathrm{\Phi}$ and $\mathrm{\Phi}'$:
    \begin{align}
        \mathrm{\Phi} * \mathrm{\Phi}'(g_k)=\sum_{\substack{p,q: \\ g_p g_q=g_k}}\mathrm{\Phi}(g_p)\mathrm{\Phi}'(g_q). \label{group_table_2}
    \end{align}
    Using Eq.~(\ref{group_table_2}) in Eq.~(\ref{group_table_1}), we finally have
\begin{align}
    a \cdot a' &= \sum_{k} g_{k} \otimes (\mathrm{\Phi} * \mathrm{\Phi}')(g_k),
\end{align}
which completes the proof.

\section{Proof of Bochner's theorem for matrix valued maps on finite group} \label{Bochner's theorem for matrix valued maps}
    Let $G$ be a finite group, and let $\Phi: G \to M_n(\mathbb{C})$ be a linear map. Recall the Fourier inversion formula [see Eq.~(\ref{eq:Fourier inversion map for group})] for maps on finite group:
    \begin{align} 
    \Phi(g)= \frac{1}{|G|} \sum_{\rho \in \mathrm{Irr}(G)} d_{\rho} \tr_{d_{\rho}} \big[(\rho (g^{-1}) \otimes \mathbb{I}) \widehat{\Phi}(\rho) \big].
\end{align}
Consequently,
\begin{align}
    \Phi(g^{-1}g')= \frac{1}{|G|} \sum_{\rho \in \mathrm{Irr}(G)} d_{\rho} \tr_{d_{\rho}} \big[(\rho (g'^{-1}g) \otimes \mathbb{I}) \widehat{\Phi}(\rho) \big].
\end{align}
Now, using the above equation, we compute
\begin{align}
    &\sum_{g,g' \in G} \overline{\inner{\Phi(g^{-1}g')h_{g'},h_g}}\nonumber\\
    &= \sum_{g,g' \in G}\frac{1}{|G|} \sum_{\rho \in \mathrm{Irr}(G)} d_{\rho}\overline{ \inner{ \tr_{d_{\rho}} \big[(\rho (g'^{-1}g) \otimes \mathbb{I}) \widehat{\Phi}(\rho) \big]h_{g'},h_g}} \nonumber\\
    &= \sum_{g,g' \in G}\frac{1}{|G|} \sum_{\rho \in \mathrm{Irr}(G)} d_{\rho} \inner{h_g, \tr_{d_{\rho}} \big[(\rho (g'^{-1}g) \otimes \mathbb{I}) \widehat{\Phi}(\rho) \big]h_{g'}} \nonumber\\
    &= \sum_{g,g' \in G}\frac{1}{|G|} \sum_{\rho \in \mathrm{Irr}(G)} d_{\rho} \mathrm{tr}\left(\tr_{d_{\rho}} \big[(\rho (g'^{-1}g) \otimes \mathbb{I}) \widehat{\Phi}(\rho) \big]h_{g'}h_g^{\dagger}\right) \nonumber\\
    &= \sum_{g,g' \in G}\frac{1}{|G|} \sum_{\rho \in \mathrm{Irr}(G)} d_{\rho} \mathrm{tr}\left((\rho (g')^{\dagger}\rho(g)) \otimes h_{g'}h_g^{\dagger}) \widehat{\Phi}(\rho)\right) \nonumber\\
    &= \frac{1}{|G|} \sum_{\rho \in \mathrm{Irr}(G)} d_{\rho} \mathrm{tr}\left(\Bigg(\sum_{g' \in G}\rho (g')^{\dagger}\otimes h_{g'}\Bigg)\Bigg(\sum_{g \in G}\rho(g) \otimes h_g^{\dagger}\Bigg) \widehat{\Phi}(\rho)\right) \nonumber\\
     &= \frac{1}{|G|} \sum_{\rho \in \mathrm{Irr}(G)} d_{\rho} \mathrm{tr}\left(A(\rho,h)^{\dagger}A(\rho,h) \widehat{\Phi}(\rho)\right), \label{SS13}
\end{align}
where $$A(\rho,h):=\sum_{g \in G} \rho(g) \otimes h_g^{\dagger}$$ and $h:=\{h_g \}_{g \in G} \subset \mathbb{C}^n$. 
\\

Now, if we assume that $\widehat{\Phi}(\rho)$ is positive semidefinite for all $\rho \in \rm{IrrU}(G)$, then from Eq.~(\ref{SS13}) we have $\sum_{g,g' \in G} \overline{\inner{\Phi(g^{-1}g')h_{g'},h_g}} \geq 0$ for all subsets $h\subset \mathbb{C}^n$, which implies that $\Phi$ is a positive definite map. This completes the proof in one direction.
\\

For the other direction, let us fix $\rho^0 \in \rm{IrrU}(G)$ and choose $h^{0}$ as the subset consisting of the vectors $h^0_g$ of the form
\begin{align}
h^{0}_g= \tr_{d_{\rho^0 }} \left(\rho^0(g) \otimes \mathbb{I}) B^{\dagger}\right),
\end{align} 
where $B \in M_{d_{\rho^0}}(\mathbb{C}) \otimes \mathbb{C}^{n*}$, and $\tr_{d_{\rho^0}}$ denote the trace over the space $M_{d_{\rho^0}}(\mathbb{C})$. Now, using Schur orthogonality relations,
we have 
\begin{align}
    A(\rho,h^0)&=\begin{cases}
       \frac{|G|}{d_{\rho^0}} B & \text{when}~\rho=\rho^0, \\
       0 & \text{when}~\rho \neq \rho^0.
    \end{cases}
\end{align}
Consequently, substituting the above form of $A(\rho,h^0)$ in Eq.~(\ref{SS13}), we obtain 
\begin{align}
    \sum_{g,g' \in G} \overline{\inner{\Phi(g^{-1}g')h^0_{g'},h^0_g}}
     &= \frac{d_{\rho^0}}{|G|}  \mathrm{tr}\left(A(\rho^0,h^0)^{\dagger}A(\rho^0,h^0) \widehat{\Phi}(\rho)\right) \\
     &= \frac{|G|}{d_{\rho^0}}  \mathrm{tr}\left(B^{\dagger}B \widehat{\Phi}(\rho)\right).
     \label{SS14}
\end{align}
Now if we assume that $\Phi: G \to M_n(\mathbb{C})$ is positive definite, then the left hand side of the above equation is non-negative. Since $B$ and $\rho^0$ are chosen arbitrarily, $\widehat{\Phi}(\rho)$ is positive semidefinite for all $\rho \in \rm{IrrU}(G)$. This completes the proof.

\backmatter

%\bmhead{Supplementary information}

%If your article has accompanying supplementary file/s please state so here. 

%Authors reporting data from electrophoretic gels and blots should supply the full unprocessed scans for key as part of their Supplementary information. This may be requested by the editorial team/s if it is missing.

%Please refer to Journal-level guidance for any specific requirements.

%\bmhead{Acknowledgements}

%Acknowledgements are not compulsory. Where included they should be brief. Grant or contribution numbers may be acknowledged.

%Please refer to Journal-level guidance for any specific requirements.

%\noindent
%If any of the sections are not relevant to your manuscript, please include the heading and write `Not applicable' for that section. 

%\bibliographystyle{sn-mathphys-num}
\bibliography{sn-bibliography}
%\bibliographystyle{amsplain}
%\bibliography{ref.bib}

\end{document}